%% file: paper.tex
\documentclass{egpubl}
\usepackage{egsgp19}

 \SpecialIssuePaper         
 \electronicVersion 

\ifpdf \usepackage[pdftex]{graphicx} \pdfcompresslevel=9
\else \usepackage[dvips]{graphicx} \fi

\PrintedOrElectronic

\usepackage{t1enc,dfadobe}

\usepackage{egweblnk}
\usepackage{cite}
\usepackage{amsgen,amsmath,amstext,amsbsy,amssymb,amsopn,amssymb}

\newcommand{\ie}{{\em i.e.,}}

\newcommand{\etal}{{\em et~al.}}
\newcommand{\red}[1]{\textcolor{black}{#1}}
\usepackage{setspace}
\usepackage{mathtools}
\usepackage{multirow}
\usepackage{amsfonts}
\usepackage{booktabs}

\usepackage[ruled]{algorithm2e} 

\SetAlFnt{\small}
\SetAlCapFnt{\small}
\SetAlCapNameFnt{\small}
\SetAlCapHSkip{0pt}

\title[Structural Design Using Laplacian Shells]%
      {Structural Design Using Laplacian Shells}

\author[E. Ulu, J. Mc{C}ann \& L. B. Kara]
{\parbox{\textwidth}{\centering E. Ulu$^{1}$,~J. Mc{C}ann$^{2}$
        and L. B. Kara$^{2}$
        }
        \\
{\parbox{\textwidth}{\centering $^1$Palo Alto Research Center, USA\\
         $^2$Carnegie Mellon University, USA
       }
}
}

\begin{document}

\teaser{
 \includegraphics[width=0.9\linewidth]{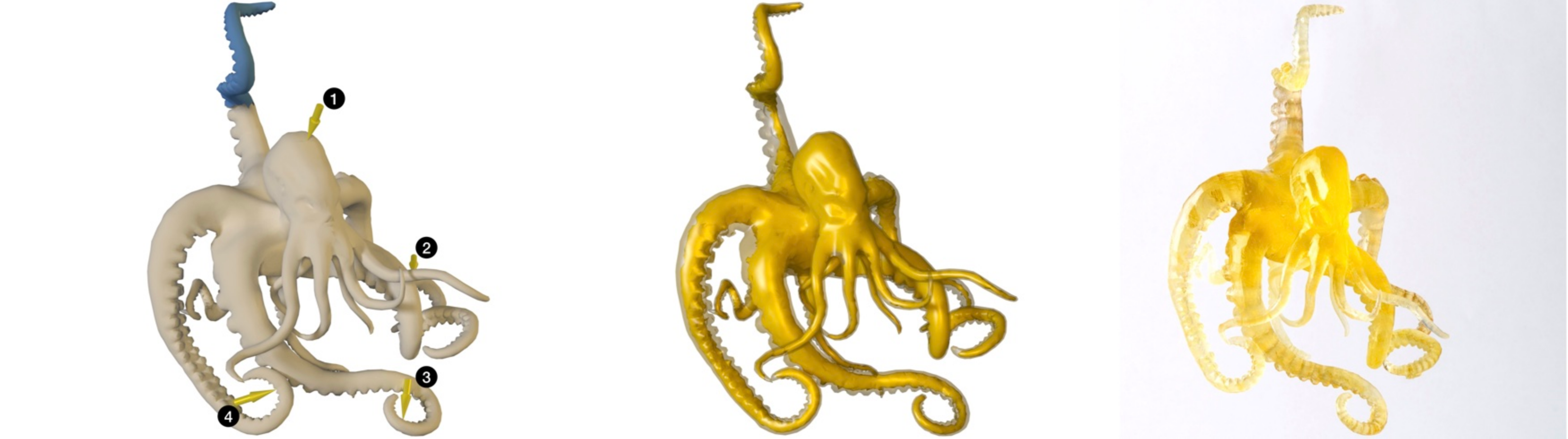}
 \centering
  \caption{We present a method for designing lightweight shell structures that are durable under the external forces that the objects may experience during their use. For a given surface mesh and a description of the possible use cases defined by the boundary conditions (blue) and external force configurations (left), our algorithm alters the shell thickness locally such that the final design (middle) can withstand external forces for any of the prescribed problem configurations. The yellow surface represents the inner boundary of the optimum shell. The optimum structure is 3D printed using a clear build material (right). The yellow soluble support structure is left inside the object to reveal the interior boundary of the shell.}
\label{fig:teaser}
}

\maketitle

\input{EGauthorGuidelines-body_with_teaser}

\bibliographystyle{eg-alpha-doi}

\bibliography{bibliography}

\end{document}

%% file: EGauthorGuidelines-body_with_teaser.tex
\begin{abstract}
   We introduce a method to design lightweight shell objects that are structurally robust under the external forces they may experience during use. Given an input 3D model and a general description of the external forces, our algorithm generates a structurally-sound minimum weight shell object. Our approach works by altering the local shell thickness repeatedly based on the stresses that develop inside the object. A key issue in shell design is that large thickness values might result in self-intersections on the inner boundary creating a significant computational challenge during  optimization. To address this, we propose a shape parametrization based on the solution to the Laplace's equation that guarantees smooth and intersection-free shell boundaries. Combined with our gradient-free optimization algorithm, our method provides a practical solution to the structural design of hollow objects with a single inner cavity. We demonstrate our method on a variety of problems with arbitrary 3D models under complex force configurations and validate its performance with physical experiments. \\

\ccsdesc[500]{Computing methodologies~Shape analysis}
\ccsdesc[300]{Computing methodologies~Mesh models}
\ccsdesc[300]{Applied computing~Computer-aided design}

\printccsdesc   
\end{abstract}  
\section{Introduction}

\begin{figure}
\centering
\includegraphics[width = \columnwidth]{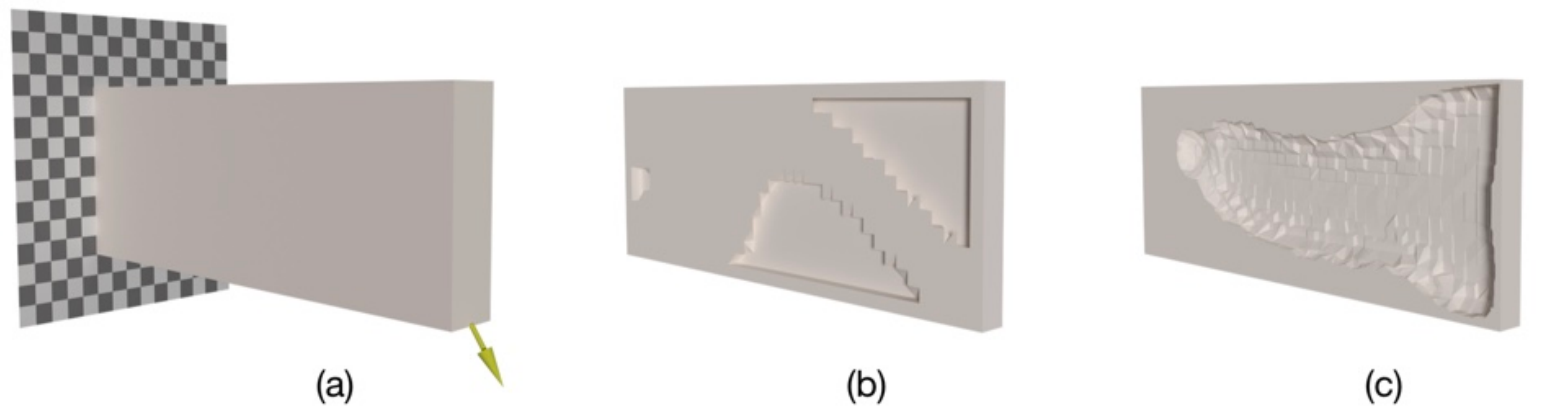}
\caption{Conventional topology optimization algorithms create internal structures disrupting the inner cavity. Shell structures can avoid this problem by adjusting the thickness locally. For a cantilever beam problem (a), example inner structures are shown for (b) conventional topology optimization and (c) shell optimization methods. Note that (b) and (c) are mid-section views.}
\label{fig:TopOptVSShellOp}
\end{figure}

As additively manufactured (AM) parts find their way into industrial applications as functional components, lightweighting and structural optimization methods have become prevalent in shape design \cite{wang2013cost,lu2014build,christiansen2015combined,langlois2016stochastic,zhou2016direct}. In many such methods, the resulting shape usually contains complex internal structures, even for very simple loading configurations. These structures create multiple isolated volumes inside the object and thus, result in manufacturability as well as functionality problems (Figure~\ref{fig:TopOptVSShellOp}). They prevent easy removal of material encapsulated inside these cavities such as internal supports in fused-deposition modeling (FDM) or excess material in selective laser sintering (SLS) and stereolithography (SLA). The object needs to be cut into pieces or perforated with multiple holes to access each cavity during post-processing to clear the encapsulated material \cite{musialski2015reduced}. However, when extensive, these destructive processes may significantly degrade the mechanical performance and invalidate the structure completely. Furthermore, complex internal structures disrupt the inner cavity that could be used for functional purposes such as housing electronics, mechanisms or wiring \cite{gao2015revomaker,song2017computational}. Shell structures provide a compelling alternative where the object is interpreted as the solid enclosed between two 2-manifold surfaces creating a single connected cavity inside. 

We propose a new structural optimization approach  for designing minimum weight shell objects. Our approach takes as input (1) a 3D shape represented by its boundary surface mesh and (2) description of the external forces that the object may experience during its use, and produces a minimum weight 3D shell structure that withstands any of the prescribed force configurations (Figure~\ref{fig:teaser}). 

For shell structure design, a reasonable approach would be to compute the optimal structure by offsetting the original boundary surface inwards in the normal direction to create the inner boundary. However, this strategy fails to guarantee that the resulting offset surface is free of self-intersections. Even advanced approaches that aim to minimize such self-intersections through local alterations of the offset directions  \cite{musialski2015reduced,zhao2017stress} do not guarantee self-intersection free offset surfaces for large thickness variations in high-curvature regions. Additionally, such an approach requires each candidate shell structure to be remeshed to perform finite element analysis (FEA) for evaluating the structural performance under the prescribed force configurations. 

Our approach overcomes these challenges using a \emph{temperature field}--a 3D scalar field obtained by solving the steady-state heat conduction--which serves as a proxy to the varying shell thickness within the shape boundary. This temperature field is the solution to the Laplace equation and the resulting offset surface is by design intersection-free even for large thickness variations. Additionally, the temperature field defined on a constant volumetric mesh allows easy and accurate transition to material distribution within the current shape hypothesis. This capability enables each step of the shape optimization to perform FEAs without requiring costly resmeshing operations. Our approach addresses classical structural design problems with fixed and known external forces \cite{andreassen2011efficient} as well as the more general class of problems ranging from multiple external force cases \cite{james2009structural} to force location uncertainties \cite{ulu2017lightweight}. Driven by a gradient-free optimization algorithm, our method provides a practical solution for designing robust hollow structures with a single inner cavity while preserving the appearance of the input model.

The main contributions of the proposed work are:
\begin{itemize}
\item a novel formulation for shell structure design involving structural mechanics,
\item a heat-based shape parametrization method that allows large variations in thickness while guaranteeing self-intersection-free boundaries in the resulting structure,
\item a gradient-free shape optimization approach to arbitrary 3D problems with complex force configurations including multiple loads as well as uncertainties in force locations.
\end{itemize}

\section{Related Work}
Our review is comprised of the studies that focus on \emph{design for fabrication}, \emph{structural analysis}, and \emph{shell object synthesis}. We emphasize approaches involving structural optimization for additive fabrication.  

\paragraph*{Design for fabrication}
A large body of work has investigated optimization techniques offering design aids to create shapes that meet the prescribed structural objectives and fabrication constraints \cite{livesu2017from}. Recent examples include designing for deformation behavior \cite{bickel2010design, skouras2013computational, panetta2015elastic, ma2017computational}, meta-materials \cite{zehnder2017metasilicone, martinez2018polyhedral, ion2016metamaterial, zhu2017two}, and lightweighting \cite{bendsoe1989optimal, lu2014build, wang2017cross}. Broader methods that can handle variety of requirements have also been investigated in \cite{chen2013spec2fab, christiansen2015combined, musialski2016nonlinear, schulz2017interactive}. Our problem falls under the category of optimum weight structure design subject to external forces, however our approach aims to generate a specific family of geometries -- shell objects.  

Other design approaches has been explored to generate models that do not require internal support structures to fabricate them. Topology optimization methods \cite{langelaar2016topology, allaire2017structural}, slice-based hollowing methods \cite{wang2018support} and specialized infill structures \cite{wu2016self, lee2018support} are investigated. Although these methods are well-suited for FDM or SLA printing processes, resulting complex structures with large number of disconnected cavities are not desirable for processes such as SLS or polyjet where excess material is typically unavoidable. Our shape parametrization is complementary to these approaches in that the presented support constraints may be facilitated in our optimization to generate shell models requiring minimum or possibly no internal supports to fabricate, thereby making it practical for all of the above mentioned fabrication processes.

\begin{figure*}
\centering
\includegraphics[width = 1.0\textwidth]{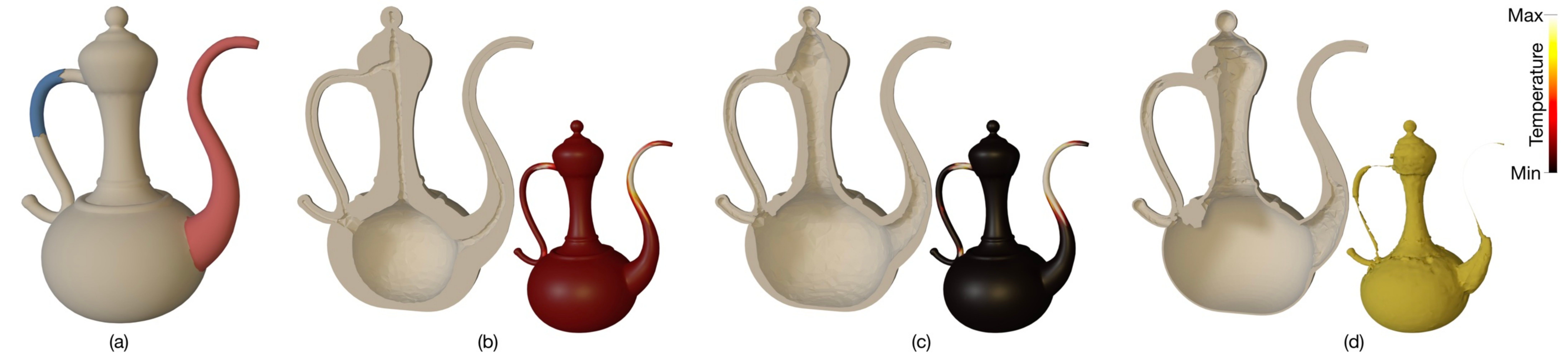}
\caption{Given a contact region (red in a) \red{where arbitrary forces can be applied on}, our algorithm optimizes the boundary thickness locally (b-c) to find the smallest weight shell structure (d) that can withstand all possible force configurations. In (b-c), we show the material distribution in two steps of the optimization. Inset figures illustrate the scalar temperature fields on the boundary that we use to drive the shell thickness in (b-c) and the removed material in (d).}
\label{fig:Overview}
\end{figure*} 

\paragraph*{Structural analysis}
In design optimization problems involving structural mechanics, structural soundness of a candidate design is determined by stress and deformation analysis through FEA \cite{ding1986shape}. Simple, low-cost elemental structures such as trusses \cite{smith2002creating, rosen2007design, kovacs2018trussformer} and beams \cite{wang2013cost, jiang2017design, wang20184DMesh} as well as higher degree of freedom elements such as tetrahedral \cite{lu2014build, schulz2017interactive} and hexahedral \cite{bendsoe2003topology, holmberg2013stress} elements are commonly employed based on the families of geometries in consideration. For thin-walled structures, more specialized elemental structures such as plate and shell elements are widely used \cite{bischoff2018models}. They allow modeling the bending behavior of thin features accurately with fewer finite elements. This makes them suitable for applications including realistic cloth simulation \cite{thomaszewski2006consistent, ly2018inverse} and design for sheet metal fabrication \cite{papeleux2002finite}. However, they are limited to surface-like structures with one dimension (\ie~thickness) being significantly smaller than the other two \cite{bischoff2018models}. In our approach, we use tetrahedral elements in FEA to analyze candidate shell designs. This allows our optimizer to explore in a larger design space by providing flexibility to generate shell models with wide range of thickness values. 

Zhou \etal \cite{zhou2013worst} introduce worst-case structural analysis by extending modal analysis used in dynamic systems to static problems in order to identify weak regions of the structure  that may fail under arbitrary force configurations. Langlois \etal \cite{langlois2016stochastic} present a stochastic finite element model to predict the failure probabilities of objects under scenarios where the loading is stochastic in nature (such as dropping and collisions). For problems with uncertainties in load direction and location spanning a small subspace,  \red{Schumacher} \etal \cite{schumacher2018set} uses parametric descriptions of the uncertainties to determine the worst-case load.

For scenarios where uncertainty in force contact location is large and \red{cannot} be described parametrically, Ulu \etal \cite{ulu2017lightweight} present a data-driven technique to predict the stress distribution for each possible force configuration and determine the one that creates the highest stress within the shape. Wang \etal \cite{wang2018efficient} improve this approach by incorporating a computationally tractable experimental design method to select data samples.
Our analysis uses a similar approach to deal with contact location uncertainty.

\paragraph*{Shell object synthesis}
Support structures \cite{hornus2018iterative}, moldable objects \cite{nakashima2018corecavity} and models with desired deformation behavior \cite{zhang2016data} are among the recent applications of shell design in the context of 3D fabrication. A common approach in generating such shell structures is to offset the object's surface to create the boundary of the inner cavity. For constant thickness offsets, Minkowski operations \cite{campen2010polygonal, martinez2015chained}, boolean operations of volumetric primitives~\cite{pavic2008high}, and use of particle sets \cite{meng2018efficiently} have been investigated. 

When the input is a polygonal surface, an offset surface can be generated by shifting the original vertices in the normal direction \cite{qu2003a3d}. Building upon this approach, Musialski \etal \cite{musialski2015reduced} present a method to create varying thickness offset surfaces. Their optimizer alters the local shell thickness to minimize a set of objective functions. However, their method is not streamlined for design problems involving structural mechanics; (1) self-intersections occur in areas of high curvature as well as regions where offsets get too large and (2) each candidate shell needs to be remeshed for FEA. The latter problem is addressed in \cite{musialski2016nonlinear} and \cite{zhao2017stress} by discretizing the shape using shell elements and adjusting their thicknesses without altering the volumetric mesh. While these methods are well-suited for optimization of thin-shells, the analysis accuracy suffers for large thickness values due to their selection of particular finite element type. Additionally, self-intersection problems may still persist for complex geometries with large thicknesses. Driven by similar motivations, we undertake both of these challenges in shell structure design. 

Our approach is similar to traditional topology optimization methods \cite{bendsoe2003topology, wang2003level} in that we define the material distribution on a fixed volumetric mesh. However, in order to enforce the resulting geometry to be a shell and address the self-intersection problems while doing so, we use a smooth scalar field defined on this volumetric mesh as a proxy to the varying shell thickness values.

\section{Algorithm}

A shell structure can be interpreted as a solid enclosed between two 2-manifold surfaces--an outer boundary $\mathcal{B}$ (enclosing  volume $\mathcal{R}$) and an inner boundary $\mathcal{B}_i$ (enclosing  volume $\mathcal{R}_i$) . For a given $\mathcal{B}$, our design problem aims to find an optimal inner boundary $\mathcal{B}_i$ such that the resulting shell structure $\mathcal{M} = \mathcal{R} \setminus \mathcal{R}_i$ has as low a mass as possible while remaining robust under the forces it experiences. We next describe our approach to addressing this problem.

\subsection{Overview}

Figure \ref{fig:Overview} illustrates our approach. Given an input 3D shape and prescribed boundary and loading configurations (Figure~\ref{fig:Overview}(a)), our system optimizes the shell thickness by manipulating a scalar field, \ie~temperature field, defined on the boundary of the object. At each step, governed by the boundary temperature distribution, our system computes the resulting steady-state temperature field inside the object. In this temperature field, the isosurface at a certain cut-off temperature forms the inner boundary of the shell structure where the region encapsulated between the outer and the inner boundaries is solid and the region inside the inner boundary is treated as void.  Then, for the current material distribution, our method computes the maximum stress encountered across the entire structure. Based on the stresses, optimization updates the boundary temperature distribution to minimize mass (Figure~\ref{fig:Overview}(b-c)). At the end, a minimum weight shell structure satisfying the imposed constraints is obtained (Figure~\ref{fig:Overview}(d)). Algorithm~\ref{alg:ourAlgorithm} summarizes our approach. 

\begin{figure}
\centering
\includegraphics[width = 0.7\columnwidth]{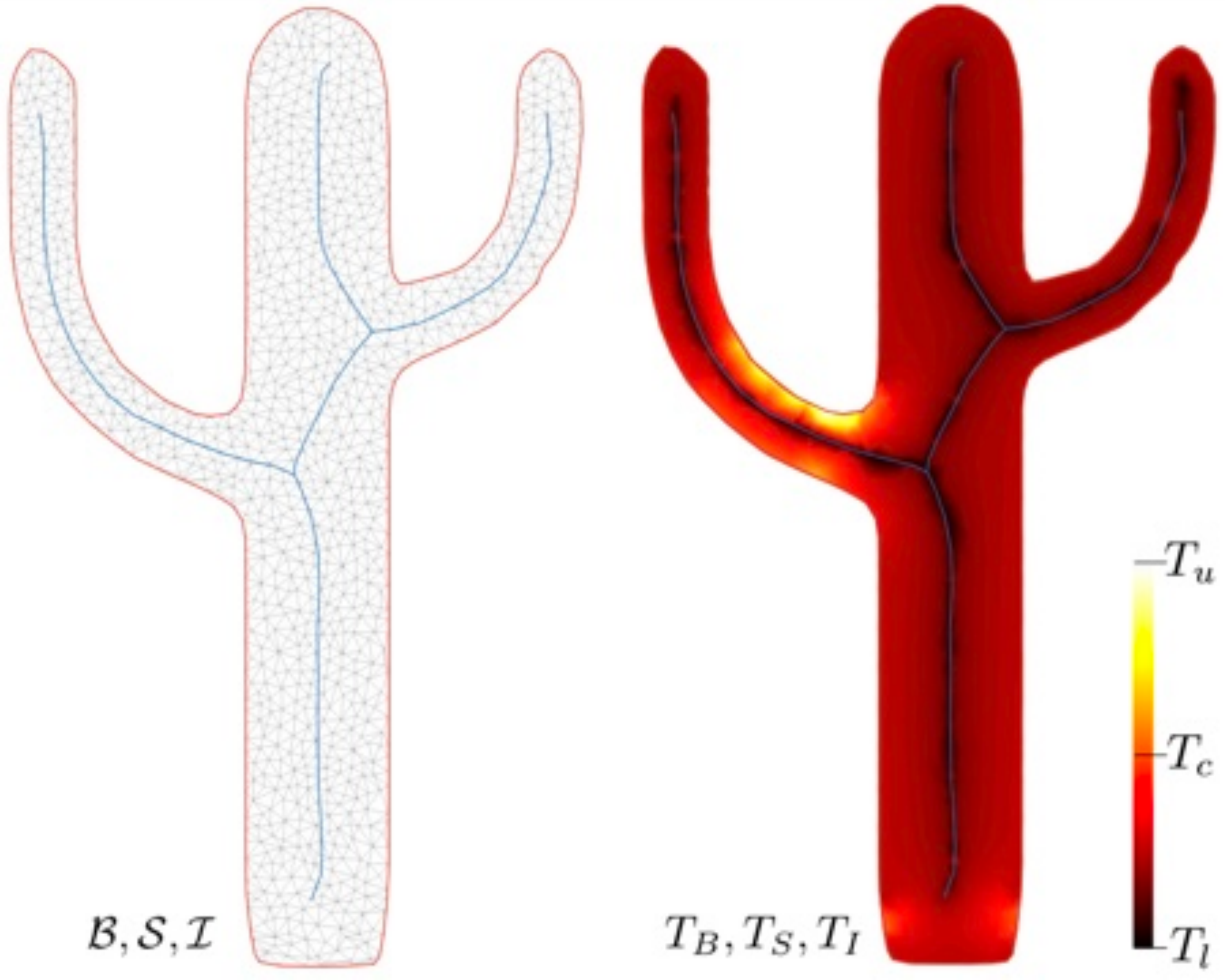}
\caption{Volumetric mesh of an object is composed of vertices on the boundary surface $\mathcal{B}$ (red), skeleton $\mathcal{S}$ (blue) and internal regions $\mathcal{I}$ (gray). For a fixed skeleton temperature, we create a temperature gradient between the boundary and the skeleton by assigning temperature values to the boundary vertices.  The isosurface at a cut-off temperature $T_c$ in the resulting temperature field constitutes the internal surface of the shell object.}
\label{fig:Parametrization}
\end{figure}

\begin{algorithm}
 \SetAlgoLined
 \SetKwInOut{Input}{Input}\SetKwInOut{Output}{Output}
 \Input{$\mathcal{B}$, boundary conditions and force configurations}
 \Output{$\mathcal{M}$}
 Initialize temperature distribution on $\mathcal{B}$, $\boldsymbol{T}_B$\;
 \While{Mass is reduced}{
  Compute the temperature distribution, $\boldsymbol{T}$\;
  Update the material distribution in $\mathcal{V}$, $\boldsymbol{\rho}$\;
  Estimate the distribution of maximum stress across all elements, $\boldsymbol{\sigma}$\;
  Compute effective boundary stress, $\boldsymbol{\tau}$\;
  Update $\boldsymbol{T}_B$\;
 }
 Extract the iso-surface $\mathcal{B}_i$ at $T_c$\;
 Construct the shell structure $\mathcal{M} = \mathcal{R} \setminus \mathcal{R}_i$\;
 \caption{Our shell structure optimization algorithm}
 \label{alg:ourAlgorithm}
\end{algorithm}

\subsection{Temperature Field}

We obtain the temperature field inside the object by creating a temperature gradient between the boundary surface $\mathcal{B}$ and the shape skeleton $\mathcal{S}$ (Figure~\ref{fig:Parametrization}). The skeleton is assigned a low temperature value and kept constant at all times while the boundary takes larger temperature values and they are adjusted to manipulate the steady state temperature field inside the object. The isosurface at a pre-defined cut-off temperature $T_c$ constitutes the inner boundary of the shell object. Here, the skeleton at the medial axis of the object serves as the inner bound for the isosurface and therefore, upper bound for the resulting shell thicknesses. We adopt the approach presented in \cite{tagliasacchi2012mean} to generate the skeleton. Due to its Laplacian-smoothing-based contraction process, this method provides smooth skeleton approximations that are less sensitive to surface details compared to other 3D skeleton generation algorithms \cite{musialski2015reduced}.  Please refer to \cite{tagliasacchi20163d} for a detailed comparison of skeleton generations methods available in the literature.

The steady-state heat distribution within a volume can be computed by solving the Laplace equation $\nabla^2 T= 0$ subject to Dirichlet boundary conditions $T|_{\mathcal{B}} = T_B$ and $T|_{\mathcal{S}} = T_S$ where $T$ is the temperature function. In the discrete setting, the linear system can be written as $\boldsymbol{\mathcal{L}} \boldsymbol{T} = \boldsymbol{0}$ where $\boldsymbol{\mathcal{L}} \in \mathbb{R}^{n_v \times n_v}$ is the discrete laplacian of the volumetric mesh $\mathcal{V}$ with $n_v$ vertices and $\boldsymbol{T}$ is the vector of per-vertex temperature values. Note that $\mathcal{V}$ is composed of the boundary vertices $b \in \mathcal{B}$,  the skeleton vertices $s \in \mathcal{S}$ and the internal vertices \red{(\ie~ Steiner points)} $in \in \mathcal{I}$, \ie~$\mathcal{V} = \mathcal{B} \cup \mathcal{S} \cup \mathcal{I}$.

Reordering the vertices as $\boldsymbol{T} = [ \boldsymbol{T}_b, \boldsymbol{T}_s, \boldsymbol{T}_{in}]'$, the Laplace equation with Dirichlet boundary conditions can be reformulated as 

\begin{equation}
\begin{bmatrix}
\boldsymbol{\mathcal{L}}_{b,b}      & \boldsymbol{\mathcal{L}}_{b,s} & \boldsymbol{\mathcal{L}}_{b,in}\\
\boldsymbol{\mathcal{L}}_{s,b}      & \boldsymbol{\mathcal{L}}_{s,s} & \boldsymbol{\mathcal{L}}_{s,in}\\
\boldsymbol{\mathcal{L}}_{in,b}      & \boldsymbol{\mathcal{L}}_{in,s} & \boldsymbol{\mathcal{L}}_{in,in}
\end{bmatrix}
\begin{bmatrix}
    \boldsymbol{T}_b\\
    \boldsymbol{T}_s\\
    \boldsymbol{T}_{in}  
\end{bmatrix}
=
\begin{bmatrix}
\boldsymbol{\cdot}\\
\boldsymbol{\cdot}\\
    \boldsymbol{0}  
\end{bmatrix}.
\label{Eq:TemperatureField_1}
\end{equation}  

\noindent Temperature field for the internal vertices can then be computed by solving the bottom block in Equation~\eqref{Eq:TemperatureField_1} for $\boldsymbol{T}_b = \boldsymbol{T}_B$ and $\boldsymbol{T}_s = \boldsymbol{T}_S$, 


\begin{equation}
 \boldsymbol{\mathcal{L}}_{in,in} \boldsymbol{T}_{in} = - \boldsymbol{\mathcal{L}}_{in,b} \boldsymbol{T}_B - \boldsymbol{\mathcal{L}}_{in,s} \boldsymbol{T}_S.
\label{Eq:TemperatureField_2}
\end{equation}

\noindent As the volumetric mesh, the boundary vertices and the skeleton vertices are constant during the process, it is sufficient to factorize the $\boldsymbol{\mathcal{L}}_{in,in}$ once as a preprocessing step and use it to solve for new $\boldsymbol{T}_B$.


One might argue that the skeleton temperatures could also be altered to provide more flexibility in shape parametrization as the cost in computing the steady state temperature distribution in Equation~\eqref{Eq:TemperatureField_2} does not change. \red{It is technically correct that each unique Dirichlet boundary conditions defined by the combination of $\boldsymbol{T}_B$ and $\boldsymbol{T}_S$ result in a unique solution to the Laplace equation and therefore, a unique temperature distribution inside the object. However, the iso-surface at $T_c$  defining the inner boundary $\mathcal{B}_i$ of the resulting shell can be the same for different $\boldsymbol{T}_B$ and $\boldsymbol{T}_S$ combinations. To avoid such redundancy, in our approach, we keep $\boldsymbol{T}_S$ constant and modify only $\boldsymbol{T}_B$ to create candidate shell structures.}

\begin{figure}
\centering
\includegraphics[width = \columnwidth]{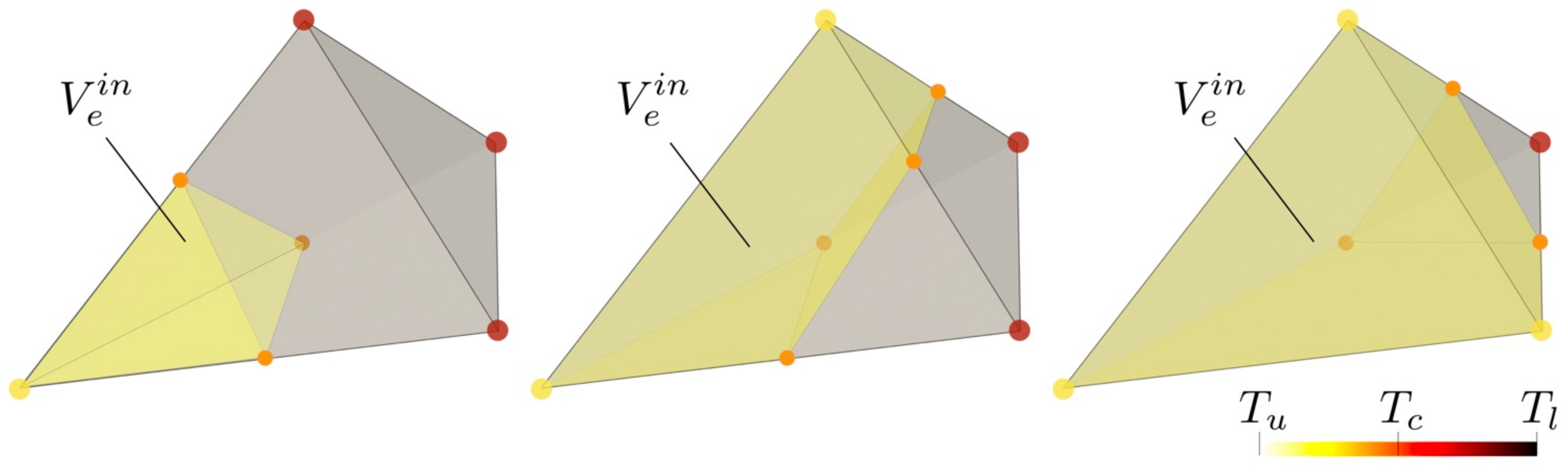}
\caption{Possible configurations that an element may take an intermediate density value. Color of a vertex indicates its temperature. }
\label{fig:DensityComputation}
\end{figure} 

\begin{figure*}
\centering
\includegraphics[width = 0.8\textwidth]{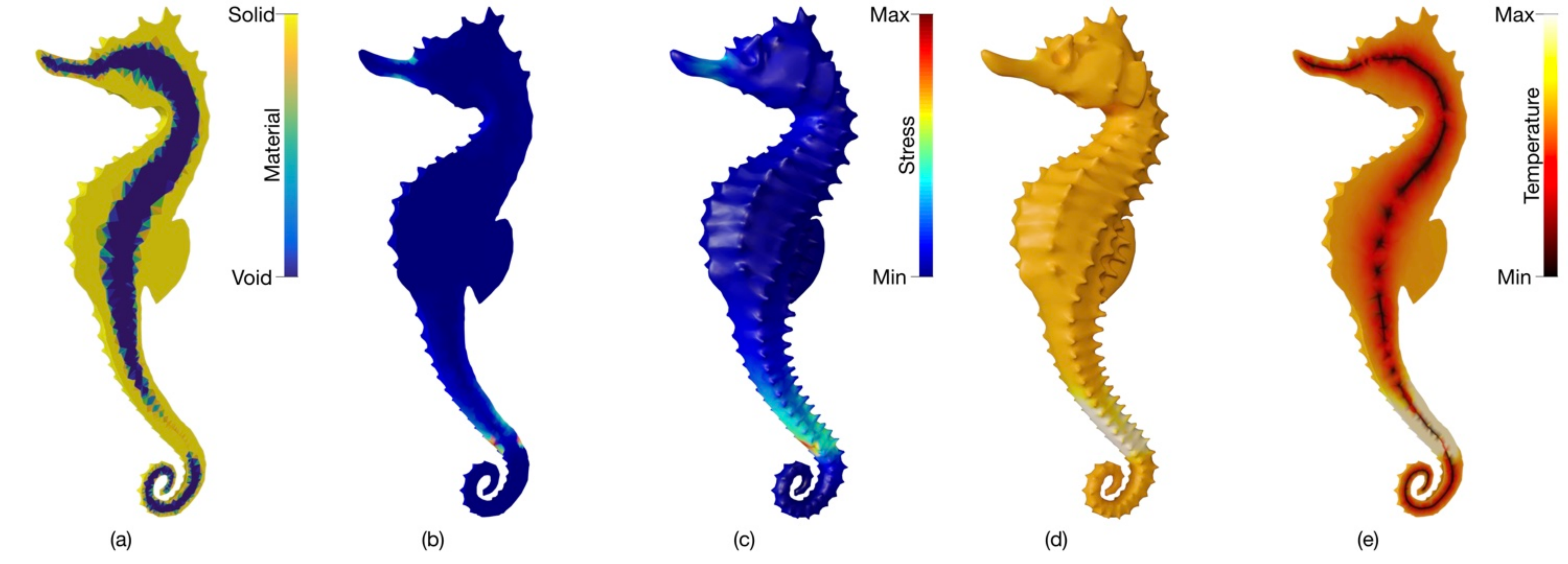}
\caption{Given the current material distribution (a), we compute maximum stresses encountered across all elements in the structure (b) and calculate their effective projection on the boundary (c). We then update the boundary temperature distribution proportional to the effective boundary stress (d) and compute the steady state temperature field inside the object (e). The isosurface created at a pre-determined cut-off temperature dictates the new material distribution for the next step of the optimization.}
\label{fig:ShapeOptimizationOverview}
\end{figure*} 

\subsection{Shape Optimization}

We tackle the following stress constrained mass minimization problem

\begin{equation}
\begin{aligned}
& \underset{\boldsymbol{T}_B}{\text{minimize}}
& & M(\boldsymbol{T}_B)  \\
& \text{subject to}
& & \boldsymbol{K}(\boldsymbol{T}_B) \boldsymbol{u}_l = \boldsymbol{f}_l \quad \forall l \in \mathcal{B}_L, \\
& & & \sigma_{cr}(\boldsymbol{T}_B) \leq \sigma_y /k ,\\
& & & T_c \leq \boldsymbol{T}_B \leq T_u
\end{aligned}
\label{Eq:optimizationProblem}
\end{equation}

\noindent where $M$ is the total mass of the solid enclosed between $\mathcal{B}$ and $\mathcal{B}_i$ created at the isosurface $T_c$ and $T_u$ is a large number defining the upper bound for the boundary temperature. $\boldsymbol{K}$ is the global stiffness matrix, $\boldsymbol{f}_l$ and $\boldsymbol{u}_l$ represent the nodal force and displacement vectors when the external force is applied to surface node $l$ \red{in a user-defined contact region $\mathcal{B}_L$}. The object fails if the critical stress $\sigma_{cr}$; maximum stress observed among all loading configurations, ever exceeds the allowable stress $\sigma_y /k$, where $k$ is the safety factor and $\sigma_y$ is the yield strength of the material. 

Finding the critical stress requires a set of FEAs--one for each distinct loading configuration. \red{As the candidate shell structure generated at each step of the optimization does not necessarily comply with the volumetric mesh $\mathcal{V}$, each candidate shell needs to be remeshed to perform FEA in the conventional form.  However, discontinuous nature of remeshing often results in computational challenges in the optimization problem. Moreover, in addition to its direct costs, remeshing introduces a computational overhead of repeated stiffness matrix computation. Namely, for each unique volumetric mesh, stiffness matrix needs to be recomputed to solve the linear elasticity problem.} We address these problems by approaching it similar to topology optimization \cite{bendsoe2003topology}, material design \cite{skouras2013computational,xu2015interactive} and microstructure design \cite{schumacher2015microstructures}. Each element in the discretized domain $\mathcal{V}$ is associated with a density variable $\rho_e$ representing whether the element $e$ is full ($\rho_e=1$) or void ($\rho_e=0$). Elements that reside between $\mathcal{B}$ and $\mathcal{B}_i$, \ie~elements with all their vertices having temperatures larger than $T_c$, are deemed to be solid while elements that are completely inside $\mathcal{B}_i$, \ie~elements with all their vertices having temperatures smaller than $T_c$, are considered to be void. For the elements that lie on the isosurface where at least one of their vertices is inside $\mathcal{B}_i$, we adopt the common approach of allowing intermediate densities $\rho_e \in [0,1]$. We determine the density of an element based on the fraction of its volume residing between $\mathcal{B}$ and $\mathcal{B}_i$. Figure~\ref{fig:DensityComputation} illustrates the possible configurations that an element is assigned an intermediate density value. For an element with volume $V_e$, the density is computed as $\rho_e = V_e^{in} / V_e$ where $V_e^{in}$ is the portion between $\mathcal{B}$ and $\mathcal{B}_i$. Assuming linear isotropic materials and small deformations, the elemental stiffness matrix $\boldsymbol{K}_e$ can be related to $\rho_e$ and the stiffness matrix for base material $\boldsymbol{K}_e^{solid}$ as

\begin{equation}
\boldsymbol{K}_e = \boldsymbol{K}_e^{void} +\rho_e^ \beta (\boldsymbol{K}_e^{solid} - \boldsymbol{K}_e^{void}).
\label{Eq:SIMP}
\end{equation}

\noindent Here, $\boldsymbol{K}_e^{void} = \epsilon \boldsymbol{K}_e^{solid}$  is the stiffness matrix assigned to the void regions to avoid singularities in FEA. We penalize the intermediate density values using a penalization factor $\beta$ \cite{bendsoe1989optimal}. We use $\epsilon = 10^{-8}$ and $\beta = 3$. For the volumetric mesh $\mathcal{V}$ with $m$ elements, one can assemble $\rho_e(\boldsymbol{T}_B)$ into vector $\boldsymbol{\rho(\boldsymbol{T}_B)} \in \mathbb{R}^m$ and construct the global stiffness matrix $\boldsymbol{K}(\boldsymbol{\rho})$ in order to determine the displacements $\boldsymbol{u}_l$ from $\boldsymbol{K} \boldsymbol{u}_l = \boldsymbol{f}_l$. Then, the stress-displacement relationship can be written as

\begin{equation}
\boldsymbol{\sigma}_l = \boldsymbol{C_g} \boldsymbol{B} \boldsymbol{u}_l, 
\label{Eq:stressFormula}
\end{equation}

\noindent where $\boldsymbol{\sigma}_l \in \mathbb{R}^{6m}$ captures the unique six elements of the elemental stress tensor
and $\boldsymbol{B}$ is the strain-displacement matrix that depends only on the elements' rest shapes. Block-diagonal matrix $\boldsymbol{C}_g \in \mathbb{R}^{6m \times 6m}$ is constructed with elemental elasticity tensors $\boldsymbol{C}_e(\boldsymbol{\rho})$ on the diagonal.
For each element, $\boldsymbol{C}_e$ can be computed analogous to $\boldsymbol{K}_e$ in Equation~\eqref{Eq:SIMP}. In our formulation, we use 10-node quadratic tetrahedral elements. Note that vertices of $\mathcal{V}$ constitute the corner nodes of the quadratic elements and additional middle nodes are generated for FEA purposes only.  

The  approach formulated in Equation~\eqref{Eq:SIMP}-\eqref{Eq:stressFormula} is useful because it preserves the same discretization throughout the optimization. Additionally, it  complies well with our heat-based shell object parametrization in Equation~\eqref{Eq:TemperatureField_1}-\eqref{Eq:TemperatureField_2} by sharing the same discretization.

We solve the optimization problem in Equation~\eqref{Eq:optimizationProblem} using a gradient-free iterative approach inspired by \cite{biyikli2015proportional}. Figure~\ref{fig:ShapeOptimizationOverview} illustrates a single iteration in our method. Given the material distribution at step $t$ of the optimization, we start by computing distribution of maximum stresses across $\mathcal{V}$, $\boldsymbol{\sigma}$ as

\begin{equation}
\sigma_i = \underset{l}{\max}\left(~\sigma_i^{vm}~\right) \quad \forall l \in \mathcal{B}_L,
\label{Eq:maxStressDistribution}
\end{equation}  

\noindent where $\sigma_i$ is the maximum stress that node $i$ experiences for all possible force configurations. Here, for each loading configuration $l$, we compute the von Mises stress at node $i$ $\sigma_i^{vm}$ by extrapolating the corresponding components of $\boldsymbol{\sigma}_l$ using the element shape functions. Note that $\max(\boldsymbol{\sigma})$ constitutes $\sigma_{cr}$ in Equation~\eqref{Eq:optimizationProblem}. We then calculate the \emph{effective boundary stress (EBS)} by projecting the stresses at the internal nodes onto the boundary nodes only. The effective boundary stress distribution allows us to estimate the boundary temperature, thereby the shell thickness for the next step $t+1$ of the optimization. Thus, the higher EBS at a vertex, the more critical it is deemed and therefore the larger temperature it is assigned.

\begin{figure}
\centering
\includegraphics[width = 0.75\columnwidth]{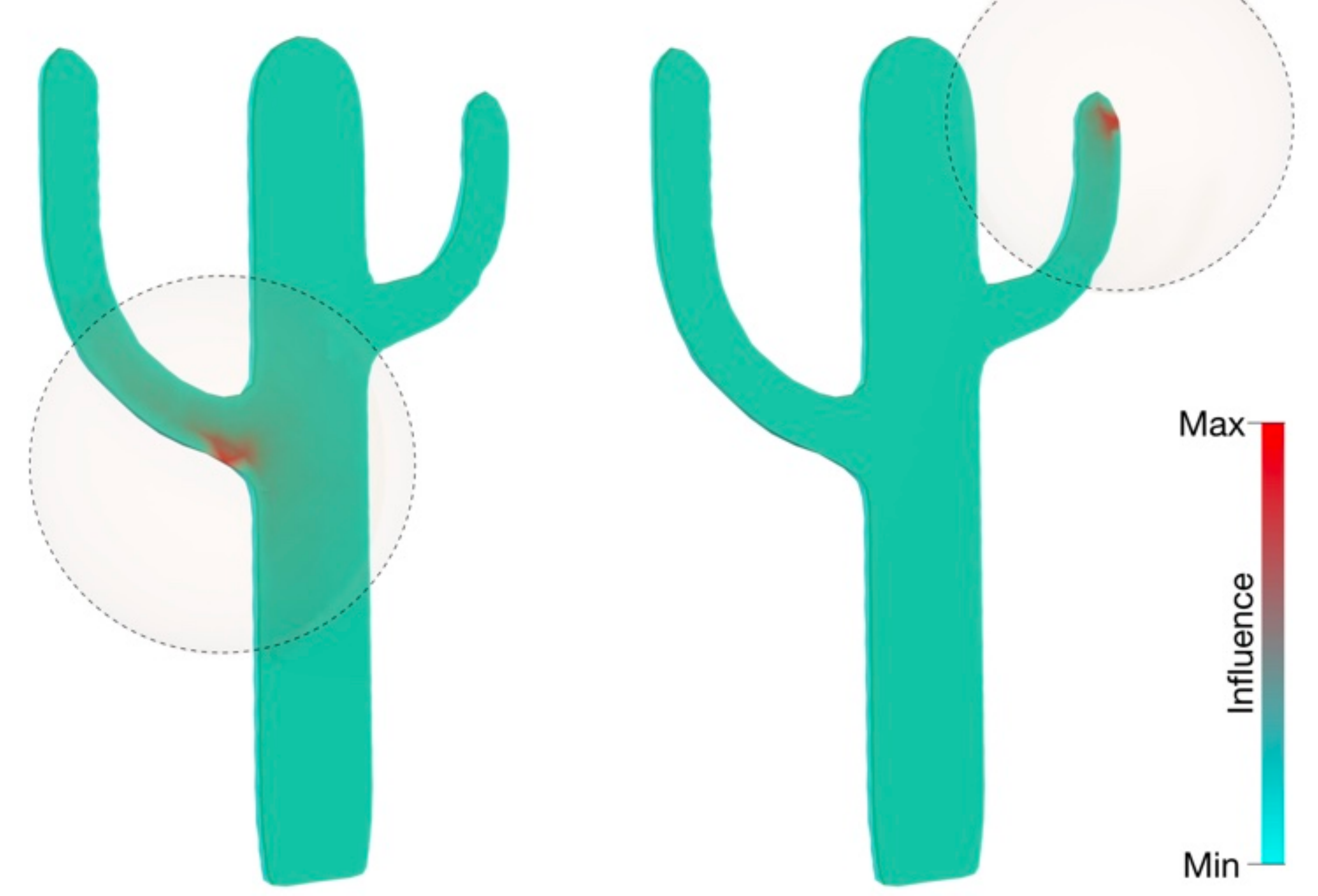}
\caption{Influence of internal nodes on boundary nodes. Circular area indicates the influence region for a boundary node at its center. The closer a node to the center, the more influence it has on the effective boundary stress at the corresponding boundary node.}
\label{fig:InfluenceCoeff}
\end{figure} 

\paragraph*{Effective Boundary Stress}
We compute the effective boundary stress by distributing the stress at internal vertices to their closest boundary vertices proportional to the distance between them. \red{The idea behind such a heuristic is based on our observation that stresses created by external forces at a certain region of a shell object can be manipulated effectively by adjusting (\ie~thickening or thinning) the shell around that particular region. Therefore, the distance-based projection allows us to adjust the boundary node temperatures and thus, the local thicknesses accordingly.} EBS at boundary vertex $j$, can be defined as 

\begin{equation}
\tau_j = \sum_{i} \sigma_i \left( \frac{1/{d_{ij}}^q}{\sum_{j} 1/{d_{ij}}^q } \right),~\forall i \in \mathcal{V} \text{ and } \forall j \in \mathcal{B}
\end{equation}
where
\begin{equation}
d_{ij} = 
\begin{cases}
dist(i,j),& \text{if }  0 \leq dist(i,j) < R, \\
\infty,& \text{otherwise}.\\
\end{cases}
\end{equation}

\noindent Here, $dist(i,j)$ is the graph distance between vertices $i$ and $j$, $R$ is the influence depth and $q$ is the influence exponent. \red{The main reason behind using graph distance as opposed to Euclidean distance here is that the graph distance takes the shape boundaries into account by constraining the paths between vertices to be within the shape.} Figure~\ref{fig:InfluenceCoeff} illustrates the contributions of internal vertices on boundary vertices for two example cases. Note that $R$ should be selected large enough that each internal node should be inside of the influence region of at least one boundary node. We use $R=10$ and $q=3$ for our examples.

\begin{algorithm}
\setstretch{1.2}
 \SetAlgoLined
 \SetKwInOut{Input}{Input}\SetKwInOut{Output}{Output}
 \Input{$^t\boldsymbol{T}_B \leftarrow \boldsymbol{T}_B$}
 \Output{$^{t+1}\boldsymbol{T}_B$}
 Scale the boundary temperature to [0,1]:\\
 \qquad ${^t\boldsymbol{T}_B}' = (^t\boldsymbol{T}_B - T_c)/(T_u - T_c)$\;
 \eIf{$\sigma_{cr}$ > $\sigma_y / k$}{
 Increase the temperature budget:\\
 \qquad $T_{\Sigma} = \sum {^t\boldsymbol{T}_B}' + h~n_b$\;}
 {
 Decrease the temperature budget:\\
 \qquad $T_{\Sigma} = \sum {^t\boldsymbol{T}_B}' - h~n_b$\;
 }
 ${^{t+1}\boldsymbol{T}_B}' \leftarrow$ Distribute $T_{\Sigma}$ to boundary vertices proportional to $\boldsymbol{\tau}$\;
 Scale back to [$T_c$, $T_u$]:\\
 \qquad $^{t+1}\boldsymbol{T}_B = {{^{t+1}\boldsymbol{T}_B}'} (T_u - T_c) + T_c$\;
  \caption{Boundary temperature update algorithm}
 \label{alg:PTO}
\end{algorithm}

\paragraph*{Boundary Temperature}
Algorithm~\ref{alg:PTO} describes how the boundary temperature $\boldsymbol{T}_B$ is updated at each step of the optimization. As the total mass $M$ of the shell object is linearly proportional to $\boldsymbol{T}_B$, the change in mass can be controlled by the change in a \textit{temperature budget} $T_{\Sigma}$; a maximum cap on the total sum of the boundary temperatures at any step of the optimization. Hence, $T_{\Sigma}$ is adjusted by a step size $h$ based on the critical stress value. $T_{\Sigma}$ is increased for $\sigma_{cr}$ larger than the allowable stress and it is reduced otherwise. The temperature budget is then distributed among the boundary vertices proportional to their corresponding EBS as

\begin{equation}
{^{t+1}\boldsymbol{T}_B}' = max \left( \frac{T_{\Sigma}}{\sum_j {\tau_j}^\kappa} \boldsymbol{\tau} ^\kappa,~1 \right).
\label{Eq:boundaryTemperature}
\end{equation}

\noindent Here, $\boldsymbol{\tau} \in \mathbb{R}^{n_b} $ is a vector storing EBS values where $n_b$ denotes the number of vertices in $\mathcal{B}$ and $\kappa$ is the proportion exponent. We use $\kappa = 5.0$ in our algorithm. 

As a final step, we linearly blend boundary temperatures from the previous step with the current step as

\begin{equation}
\boldsymbol{T}_B = \alpha~{^{t+1}\boldsymbol{T}_B} + (1-\alpha)~{^{t}\boldsymbol{T}_B},
\label{Eq:boundaryTemperature_2}
\end{equation}

\noindent where $\alpha$ is the blending factor. This helps attenuate drastic jumps between two consecutive iterations that could be created due to local stress concentrations and result in smooth transitions throughout the optimization. We found out that $\alpha=0.5$ works well for all our examples. 

As the optimization approaches toward the final result, we observe frequent direction changes in $T_{\Sigma}$  across consecutive iterations. This usually results in back and forth jumps between two states. We address this problem by halving the step size in every direction change. Hence, the step size provides an effective indicator for convergence. We initialize the optimization with $h=0.1$ and terminate it when $h$ is smaller than $10^{-8}$.

\begin{figure}
\centering
\includegraphics[width = 0.8\columnwidth]{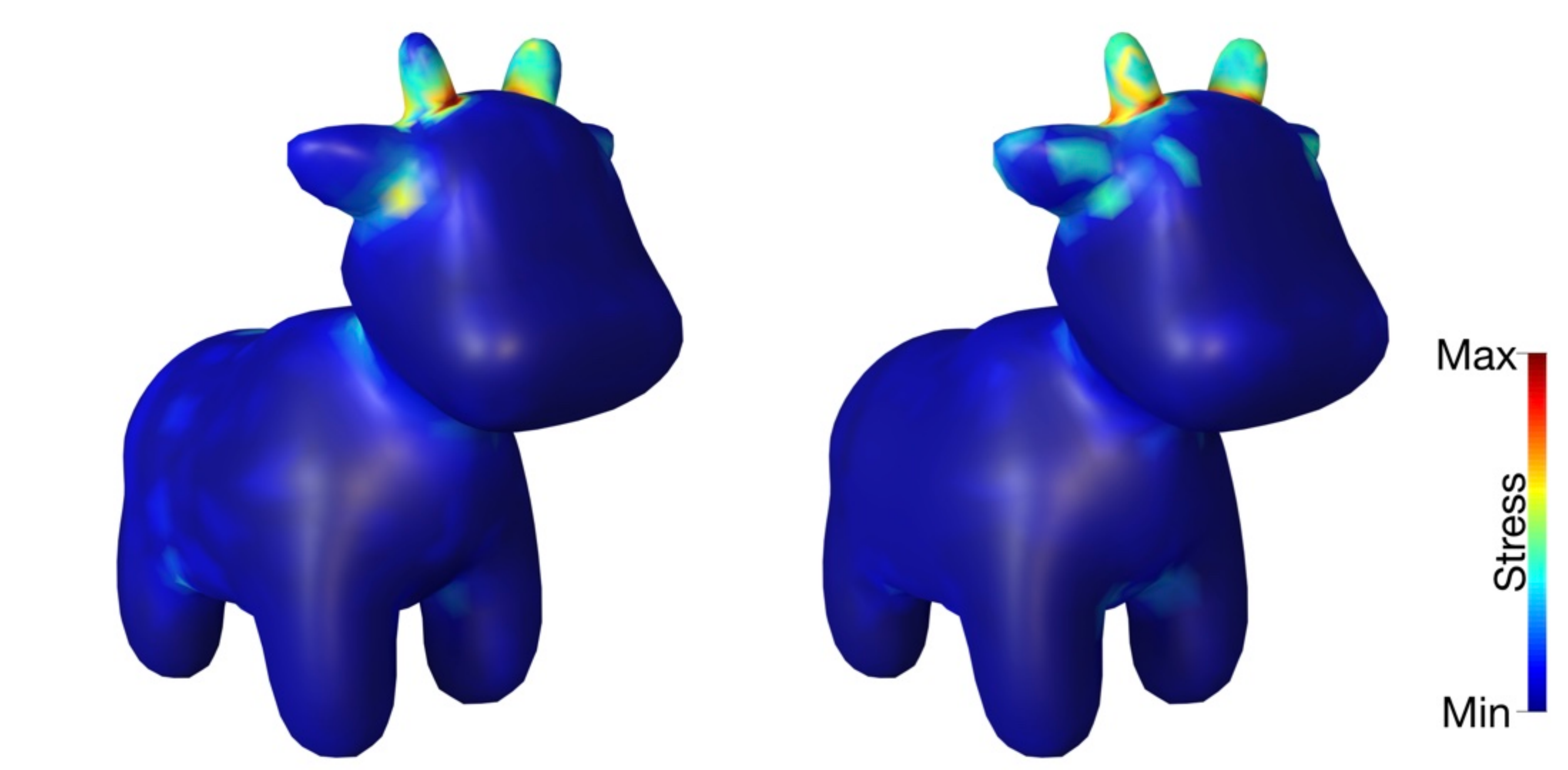}
\caption{Distribution of maximum stresses computed for a fully solid model. Estimated stress (left) is an approximation of the ground truth (right), obtained with a lower computational cost.}
\label{fig:MaxStressBFvsEstimation}
\end{figure}

\section{Force Location Uncertainty}

In the case of a small number of force configurations, the maximum stress distribution in Equation~\eqref{Eq:maxStressDistribution} and $\sigma_{cr}$ can be calculated by performing an FEA per force configuration. However, for cases where a large number of force configurations  need to be considered, a brute force approach in this way becomes restrictively impractical. A vast majority of such problems can be generalized under force location uncertainties where the external forces' contact locations exhibit significant variations during the use of the object. 

For problems with force location uncertainties, we adopt a similar approach to \cite{ulu2017lightweight} in estimating the maximum stress distribution $\boldsymbol{\sigma} \in \mathbb{R}^{n_v}$ and $\sigma_{cr}$. We perform FEAs for only a small number of force configurations and use them to construct a mapping between the nodal forces and the resulting stress distributions and estimate the stress distributions for the remaining force configurations. 

Suppose uncertainty in the external force locations is defined such that the normal forces are allowed to make contact within a user-specified union of contact regions $\mathcal{B}_L \subseteq \mathcal{B}$. We start by uniformly sampling a number of force instants on $\mathcal{B}_L$ such that the geodesic distance between the closest samples is maximized. Here, we use approximate geodesic distances \cite{crane2013geodesics} for computational efficiency.  We assume that each force is distributed to a small circular area around the contact point to avoid stress singularities. Therefore, it can be represented as a sparse vector $\boldsymbol{f}_l'$ of size $n_b$ where each element corresponding to a boundary surface node that lies inside this circular area takes a non-zero value equal to the magnitude of the nodal force component. Von Mises stress corresponding to each force instant forms a vector $\boldsymbol{\sigma}^{vm}$ of size $n_v$. 

Using a small number of training force samples $p$ (\ie~$p \ll n_b \ll n_v$),  it is not possible to represent the relationship between these two high dimensional data using a simple mapping function. Thus, we project both the force and von Mises stress data onto lower dimensional spaces; we use the Laplacian basis for forces and the principal components for stresses. We compute the Laplacian basis functions as the first $s$ eigenvectors of the surface graph Laplacian $\boldsymbol{\mathcal{L}}_\mathcal{B} \in \mathbb{R}^{n_b \times n_b}$. In all our examples, we use  $s = 15$ eigenvectors to construct our lower dimensional Laplacian basis. For stresses, we use principal component analysis (PCA) to obtain the lower dimensional basis. A PCA on the stress data results in $(p-1)$ principal vectors.

Lower dimensional representations of the force instants and the corresponding stresses allow us to construct a simple mapping between these two spaces. We build the following quadratic regression model with L2 regularization 

\begin{equation}
\boldsymbol{\mathcal{T}} = \boldsymbol{\mathcal{F}} \boldsymbol{\mathcal{W}}
\end{equation}
where $\boldsymbol{\mathcal{T}}$ is the lower dimensional representation of nodal von Mises stresses, $\boldsymbol{\mathcal{F}}$ is a matrix storing lower dimensional force representation including the quadratic terms and $\boldsymbol{\mathcal{W}}$ is the coefficient matrix that models the quadratic map. The coefficient matrix can then be computed as 

\begin{equation}
\boldsymbol{\mathcal{W}} =  (\boldsymbol{\mathcal{F}}^T\boldsymbol{\mathcal{F}} + r\boldsymbol{I})^{-1} (\boldsymbol{\mathcal{F}}^T \boldsymbol{\mathcal{T}})
\end{equation}
 where $\boldsymbol{I}$ is identity matrix and $r$ controls  regularization. Using this map, we estimate the nodal von Mises stresses $\boldsymbol{\sigma}^{vm}$ for each force instant. This allows us to compute an estimation for $\boldsymbol{\sigma}$ using Equation~\eqref{Eq:maxStressDistribution}. Figure~\ref{fig:MaxStressBFvsEstimation} illustrates $\boldsymbol{\sigma}$ obtained through our estimation in comparison to the ground truth for an example model.   
 
Note that computed $\boldsymbol{\sigma}$ here is only an approximation of the actual values, thus cannot be used directly to obtain $\sigma_{cr}$. We adopt the hierarchical search approach described in \cite{ulu2017lightweight} to compute $\sigma_{cr}$ accurately and guarantee the structural soundness of the resulting shell under prescribed force location uncertainties.

\section{Results and Discussion}

\subsection{Shape Parametrization}

\begin{figure}
\centering
\includegraphics[width = \columnwidth]{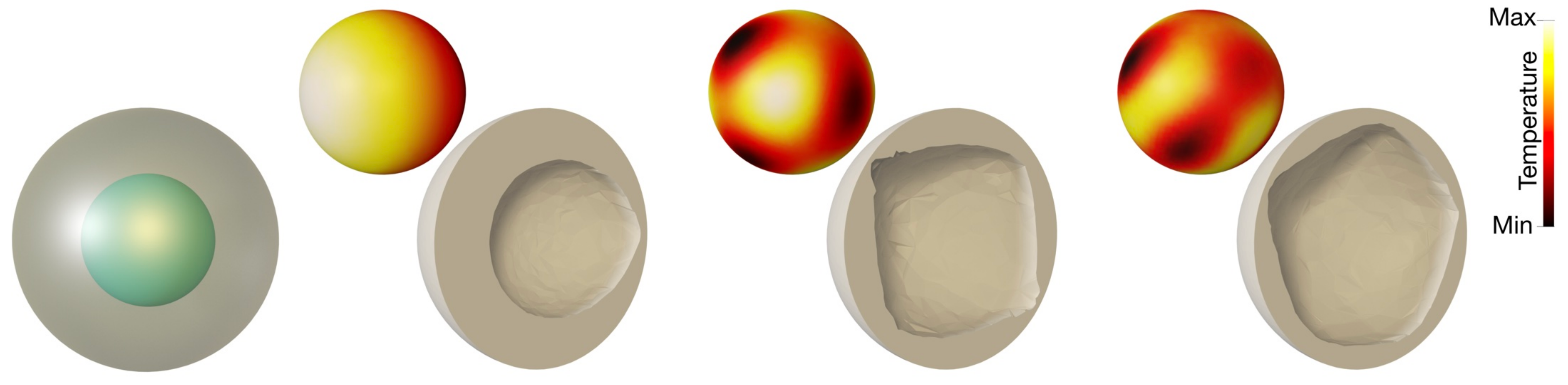}
\caption{Our temperature based shape parametrization allows  large variations in the internal surface and therefore the resulting shell structure. Example shell structures (right) obtained for a simple sphere model using a spherical skeleton (left) are illustrated. Corresponding boundary temperatures are shown on the top.}
\label{fig:ShapeParametrization}
\end{figure}

In our heat based shape parametrization, the shell structure is represented by a temperature field created on the boundary surface of the input mesh, \ie~the number of design parameters are identical to the number of boundary vertices. This provides our optimizer enough flexibility to perform localized changes with large variations on the shell thickness while preserving smoothness in the resulting internal surfaces. Figure~\ref{fig:ShapeParametrization} illustrates an example set of shells that can be achieved by our shape parametrization on a simple sphere model. Although the number of design parameters could be large for models with intricate surface details such as the octopus (Figure~\ref{fig:teaser}) or the sea horse (Figure~\ref{fig:Results-MultiForce}), our parametrization preserves a certain level of smoothness inherently due to the smooth temperature field created inside the object. 

\begin{figure}
	\includegraphics[width = 0.9\columnwidth]{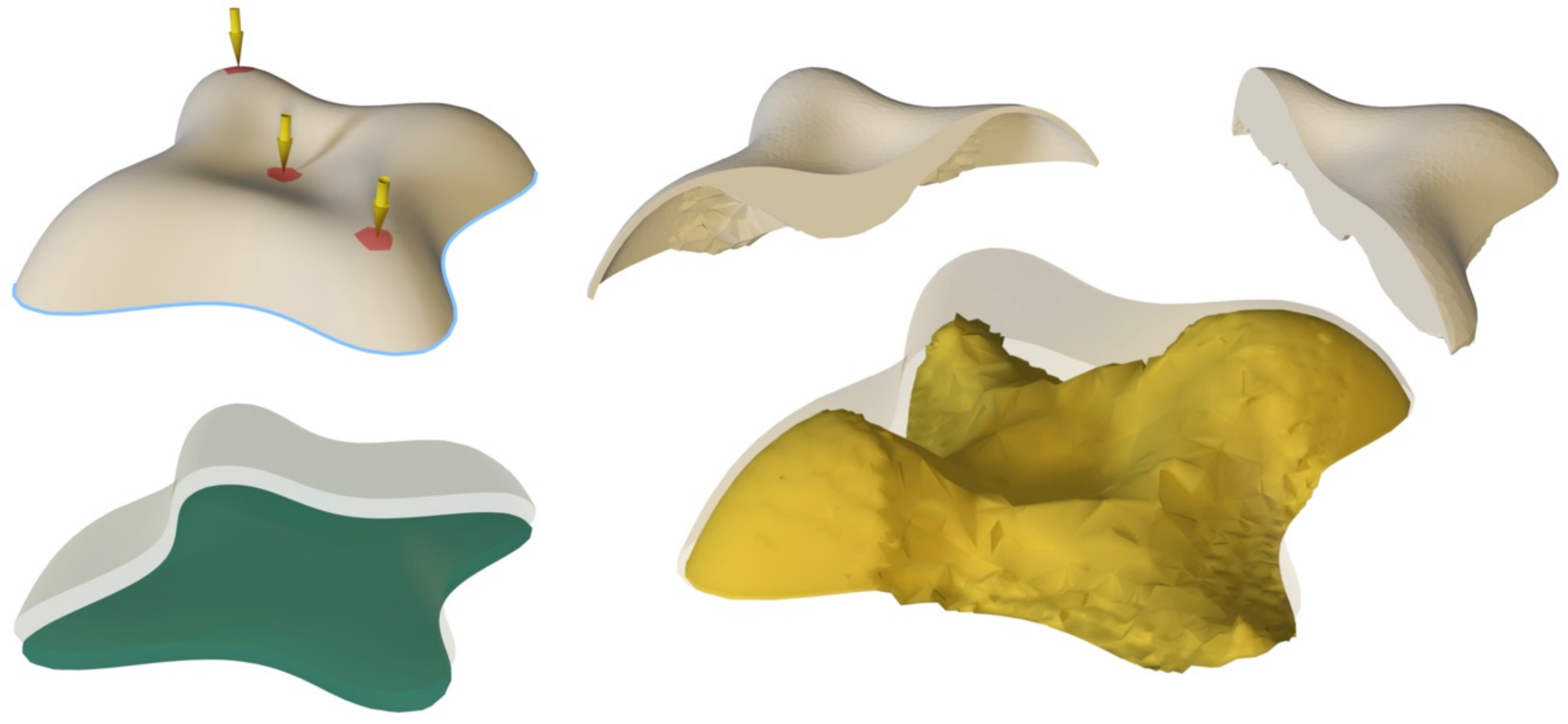}
   	\caption{An example shell optimization for open surfaces. Temperature field is generated between the input boundary surface (clear) and an external skeleton (green).  Isosurface at the cut-off temperature (yellow) constitutes the internal boundary of the resulting shell. Variation of the wall thickness is shown with diagonal cross-sections. (top-right).}
   	\label{fig:Openmesh}
\end{figure}


Another advantage of our shape parametrization method is that the resulting temperature field is easily transitioned into the density based representation from which the structural analysis can be constructed directly. This allows us to use the same volumetric mesh throughout the entire optimization without requiring costly remeshing operations. Moreover, during this transition, only a small number of intermediate density elements are created (Figure~\ref{fig:ShapeOptimizationOverview}(a)), which is highly desirable in structural optimization as the analysis accuracy is only minimally affected.

\paragraph*{Skeleton and Guarantee for a Single Hole}
Skeletons used in our examples are shown in green in Figure~\ref{fig:Results-MultiForce} and Figure~\ref{fig:Results-FLU}. Note that the skeletons are essentially a set of connected vertices in our formulation. Therefore, the variety of geometric representations including polylines, open or closed surfaces, meso-skeletons \cite{tagliasacchi2012mean} or combinations of these can be used as skeletons in our approach. In our examples, we employ curvilinear skeletons for the deer head, sea horse, cactus, spot, shark and octopus, a meso-skeleton for the mug, and an open surface mesh for the beam model.

As the skeleton defines the inner bound for the resulting shell, users may constrain the design space by manually manipulating the skeleton geometry. For example, in the pitcher model, we add a spherical surface to the curvilinear skeleton inside the inner chamber to set the minimum capacity in the final result. As the isosurface defining the inner boundary is restricted to reside between the skeleton and the outer boundary surface, the resulting structure is guaranteed to have the internal part of the skeleton hollow. This can be useful in designing dedicated housings to incorporate electronics or other instrumentation.

\begin{figure}
\centering
\includegraphics[width = \columnwidth]{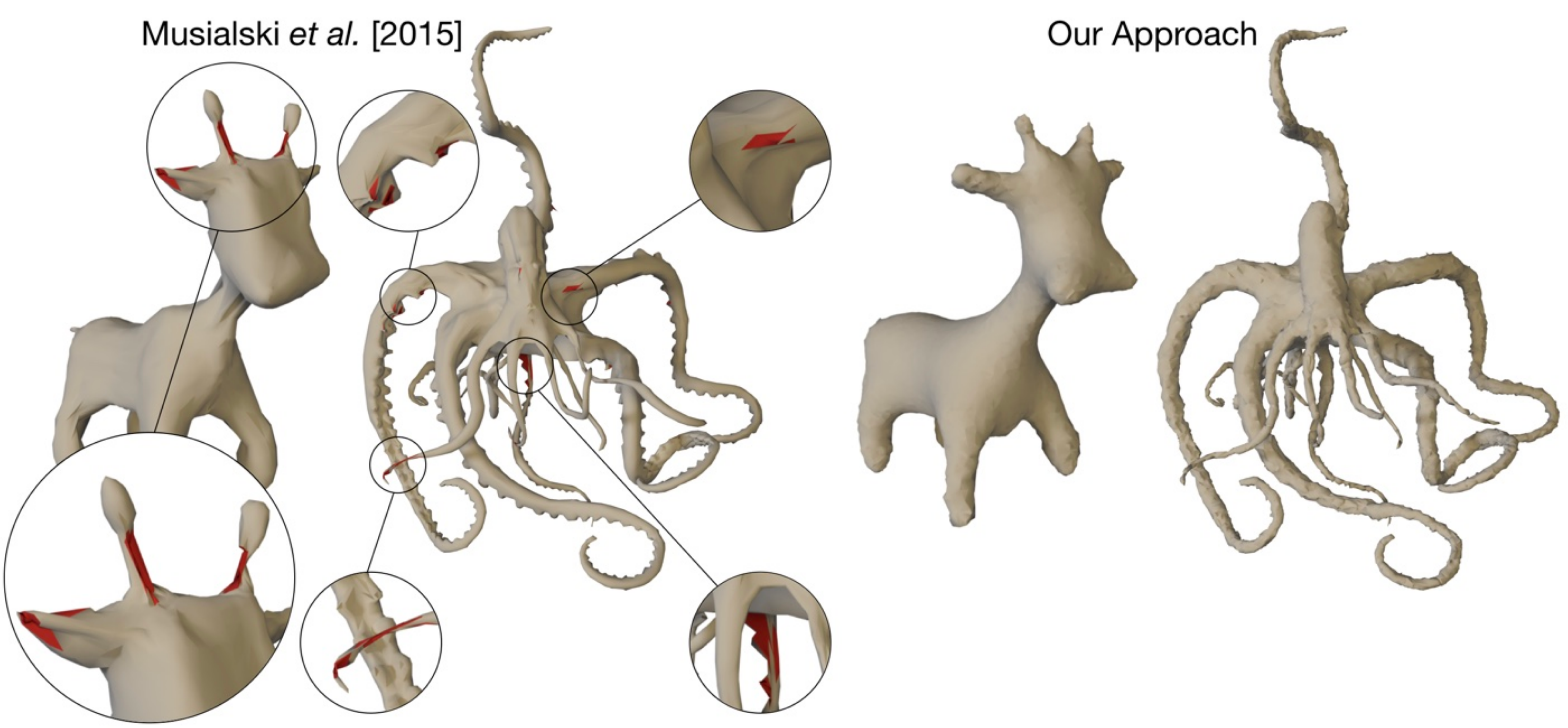}
\caption{Comparison of inner boundaries created by Musialski \etal~\cite{musialski2015reduced} and our method for random sets of design variables. Their method results in self-intersections for large offset values around high-curvature regions of the boundary surface. Our approach circumvents such challenges even for complex geometries such as the octopus. Self-intersections are shown in red.}
\label{fig:ComparisonOffsetSurfaces}
\end{figure} 

\begin{table*}
\setstretch{1.2}
\small
\caption{Performance of our algorithm on a variety of models. Columns 2 to 5 are the number of finite elements in the volumetric mesh and the number of vertices on the boundary, skeleton and remaining regions, respectively. Column 6 reports the number of force configurations for which the shell object is optimized. In all examples, the target factor of safety is $90\%$ of that of the fully solid versions.} 
\centering 
\begin{tabular}{l cccccccccc} 
\toprule
\multirow{2}{*}{Model} & \multirow{2}{*}{Elements} & \multicolumn{3}{c}{Vertices} & Force &\multirow{2}{*}{Iteration} & \multirow{2}{*}{Time [s]} & \multicolumn{3}{c}{Volume [cm\textsuperscript{3}]}\\ \cline{3-5} \cline{9-11}
& & Boundary  & Skeleton  & Internal & Configurations & & & Initial & Optimized & Reduction [\%]\\ [0.5ex]
\midrule
Cactus			& 	43727 & 1856	& 233	& 	5858 & 1 & 29	& 11.60 & 68.99	& 26.51 & 61.6\\
Sea Horse		& 	99054 & 7502 & 105	& 11814	& 1 & 42 & 157.23 &	6.42 & 1.48  & 77.0\\
Deer Head	& 	89521 & 7500 & 	834 & 8969	& 20 & 47 & 92.68 & 27.80 & 9.14 & 67.1\\
Lilium			&	41962	&	3389	&	2085 &	 4513	&	3	&	37 	&	24.20	&	-	&	91.35	& -	\\
Octopus		&	130896	 &	12502	& 	893	&	14035	& 4	&	53	&	70.69	&	75.32 &	 26.00 & 65.5	\\
Test Beam	&	87284	&	6000 &	976	&	9730	 &	 2	&	38	&	76.84	&	20.38	&	9.32	& 54.3	\\
Pitcher			& 	61881 & 4997 & 	735 & 6703	& 820 & 92	& 1788.39 	& 175.85 &	31.09 & 82.3 \\
Spot				&	56504 & 2930 & 310 &	7247 	&	2189 	&	50 &	1037.77 &	367.75 &	132.04 & 64.1\\
Mug				&	48604	&	3000 &	 2679 &	 3575	&	2510		&	76 	&	1212.17	&	104.12	&	46.26		& 55.6\\
Shark				&	69243	& 5757	&	680	&	7522 	&	4282	&	33	&	1100.37		&	66.45	& 12.79	& 	80.8\\
\bottomrule
\end{tabular}
\label{tab:Results}
\end{table*}

Importantly, because the solution to the heat equation is a harmonic function, the maximum principle \cite{axler2001harmonic} guarantees that all voids are connected to the skeleton and all solid areas are connected to the outer boundary. This means that an optimized object will never have more voids than the number of connected components in the skeleton. To ensure the creation of a single hole, we restrict the number of skeletons to one in our approach.

Although the internal skeletons can be obtained easily for closed meshes, they are not well-defined for open surfaces. Figure~\ref{fig:Openmesh} demonstrates an example case where the input to our algorithm is an open surface mesh and a solid shell structure is  to be created by \emph{thickening} the input surface. The surface is fixed at its boundary and three different loads are applied to top of the surface. In this case, we create a surface to serve as the skeleton (shown in green) in our algorithm by projecting the mesh to a lower plane. Then, the volumetric mesh is created in the space enclosed between these two surfaces, on which the temperature field is solved. The resulting shell boundary is shown in yellow. The combination of this boundary and the input surface constitutes the resulting solid shell.

\paragraph*{Comparison}
Figure~\ref{fig:ComparisonOffsetSurfaces} compares the performance of our shape parametrization with the reduced order offset surfaces approach of Musialski \etal~\cite{musialski2015reduced}. The offset directions in this method are computed such that the self-intersections are avoided as much as possible. However, high curvature regions are still susceptible to such problems for large thickness values. Therefore, there is no guarantee that the candidate shell structures generated during the optimization will be free of self-intersections. Although self-intersections might be tolerated for certain optimization problems, design optimization involving structural mechanics are highly sensitive to such complications as each candidate design needs to be evaluated by a set of FEAs. In our method, because the inner shell boundary is a level-set of a smooth and continuous temperature field, the resulting shell is guaranteed to be self-intersection free. Note that the inner boundaries shown in Figure~\ref{fig:ComparisonOffsetSurfaces} are results of random sets of feasible design variables. For the offset surfaces, we use $36$ manifold harmonics to generate the results. In both cases, we use the same skeleton structure generated using \cite{tagliasacchi2012mean}. Parts of the skeleton that are outside of the boundary are manually corrected. 

\subsection{Structural Optimization}

\subsubsection{Multiple Problem Configurations}
\begin{figure*}
\centering
\includegraphics[width = 1.4\columnwidth]{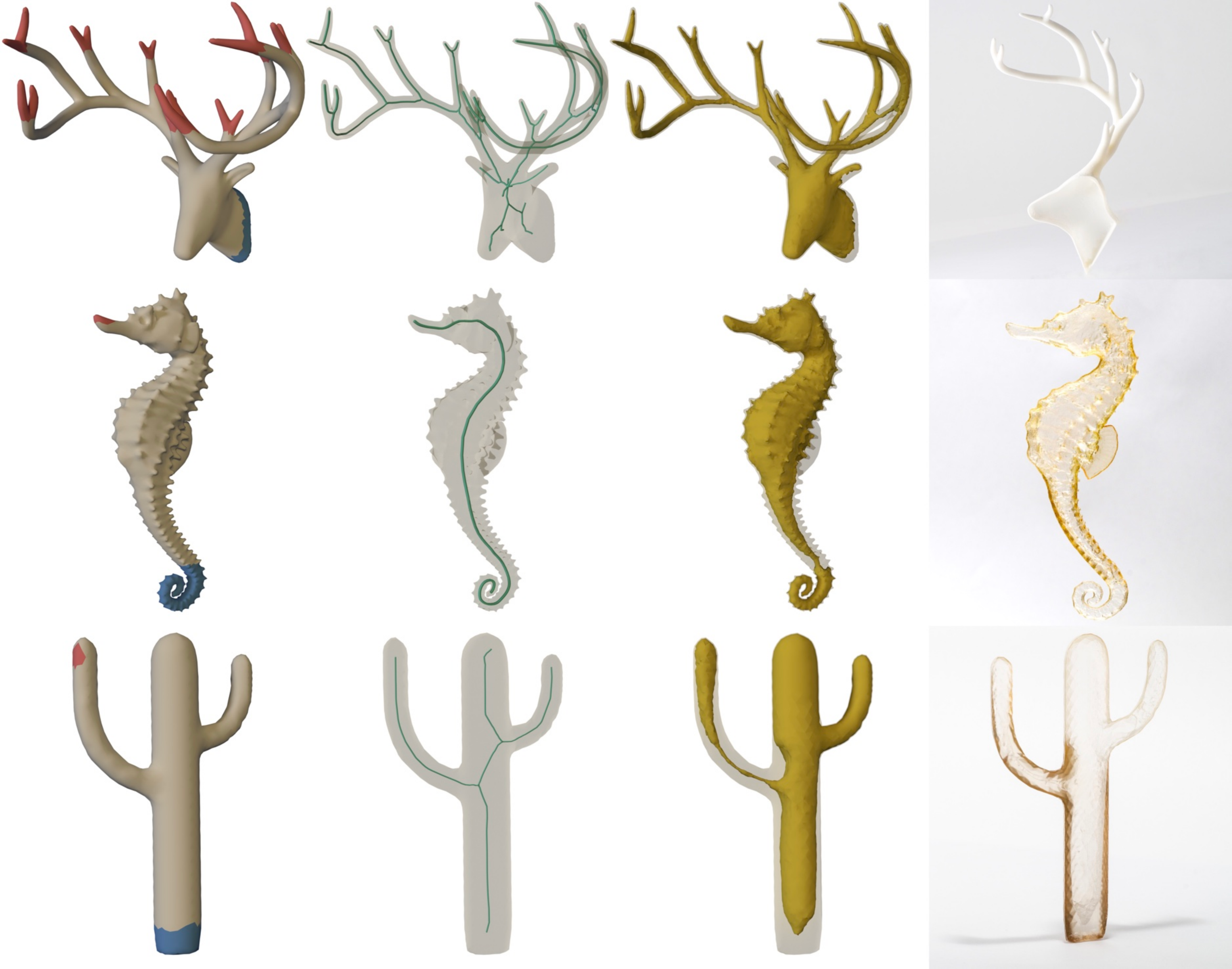}
\caption{Structural optimization results for problems with multiple force configurations. Left-to-right, problem setup with fixed boundary conditions (blue) and force contact regions (red), skeletons used during the optimization, optimized shell structures and their 3D printed cut-outs revealing the variations in the shell thickness. Yellow surface indicates the inner boundary of the shell.}
\label{fig:Results-MultiForce}
\end{figure*}

Figure~\ref{fig:Results-MultiForce} illustrates the results of our method on various 3D problems where the shell structures are required to withstand a number of fixed and known force configurations. Displacement boundary conditions are shown in blue and the force contact regions are shown in red. Our shell structure optimization algorithm detects the failure-prone parts of the objects and adjusts the local shell thickness accordingly. For example, in the deer head model, the branching regions of the antler where the stresses are high due to large bending moments are thickened while the rest of the head is left rather hollow. Similarly, in the cactus model, since the trunk is already structurally sound under a load applied to the long arm, only the region where the arm branches out from the trunk is optimized to have a large shell thickness.

\begin{figure*}
\centering
\includegraphics[width = 1.4\columnwidth]{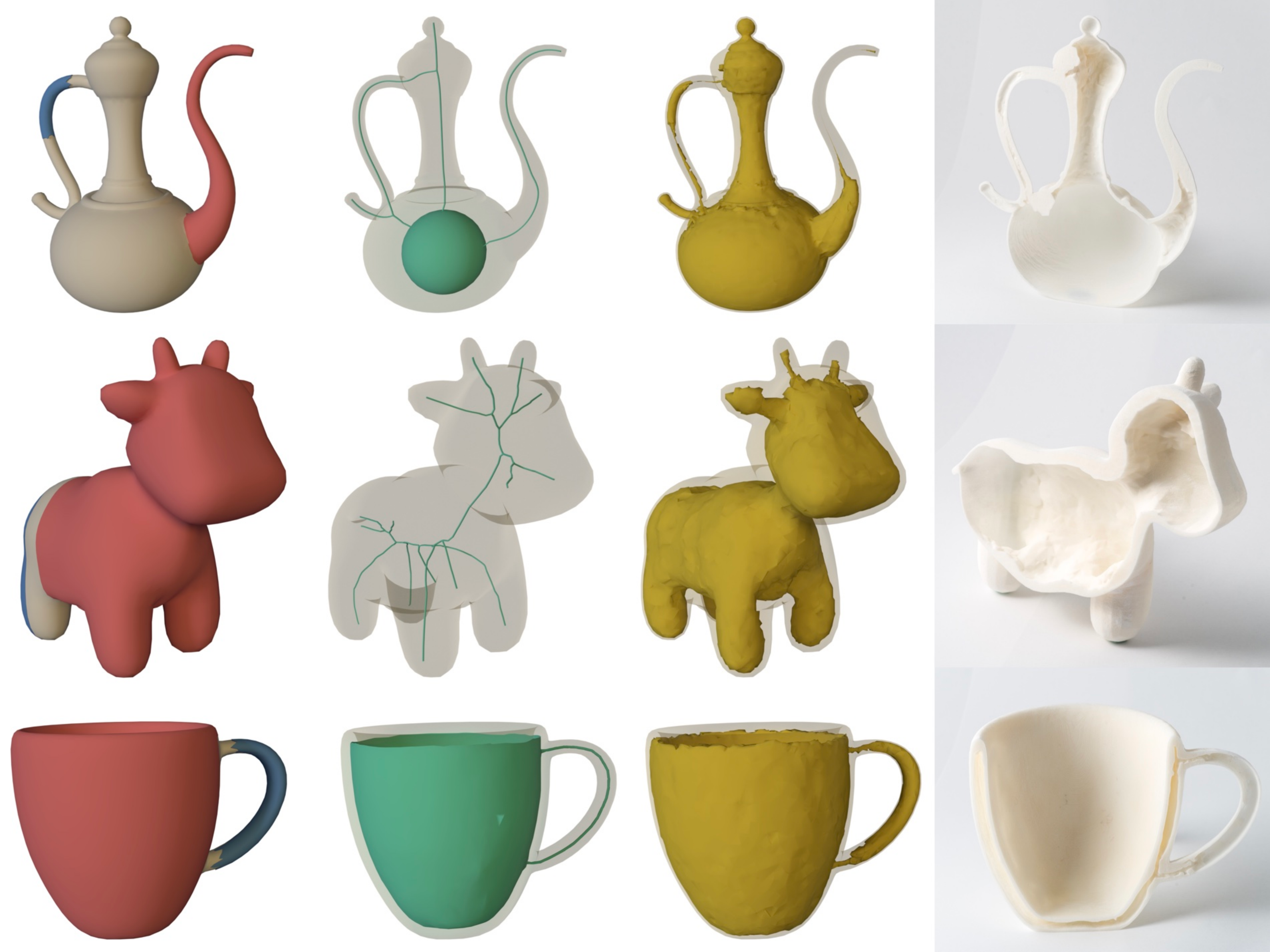}
\caption{Structural optimization results for problems with force location uncertainties. Left-to-right, problem setups with fixed boundary conditions (blue) and force contact regions (red), skeletons used during the optimization, optimized shell structures and their 3D printed cut-outs revealing the variations in the shell thickness. Yellow surface indicates the inner boundary of the shell.}
\label{fig:Results-FLU}
\end{figure*} 

Our formulation in Equation~\eqref{Eq:maxStressDistribution} also allows optimization of structures for different displacement boundary conditions as well as for any arbitrary loading configuration at a cost of increased computational complexity. In such cases, each set of Dirichlet boundary conditions needs to be incorporated into the stiffness matrix and a linear solve needs to be performed per problem configuration at each step of the optimization.  Figure~\ref{fig:MultipleBC} illustrates an example case where the beam is optimized to withstand two sets of boundary conditions as well as loading configurations. 

In all of the examples, given the boundary and loading conditions, the target factor of safety for optimization in Equation~\eqref{Eq:optimizationProblem}  is set to be $90\%$ of the factor of safety of the fully solid, original models. This effectively means the optimization tries to preserve at least $90\%$ of the structural strength of the fully solid models.  Under this relatively stringent condition, our algorithm achieves $54\%$ to $77\%$ mass reduction. Table~\ref{tab:Results} reports the reduction in volume together with various other metrics relevant to these models. 

\subsubsection{Force Location Uncertainty}
In Figure~\ref{fig:Results-FLU}, we demonstrate the results of our method on a set of example problems where there are uncertainties in the force configurations. In these examples, the displacement boundary conditions are assumed to be known and fixed and the forces are compressive and normal to the boundary surface. 

Similar to the previous cases, a thicker shell is obtained around the parts of the objects where high stresses may develop under the given force location uncertainties. Notice the handles of the pitcher and mug models or the tail connecting the shark body to the base plate. 

Although the computational cost is higher and the optimization takes longer to converge compared to the simpler loading configurations mentioned earlier, a similar volume reduction performance is obtained. For the problems with force location uncertainties, we achieved $55\%$ to $82\%$ reduction in volume (Table~\ref{tab:Results}).

\begin{figure}
\centering
\includegraphics[width = 0.9\columnwidth]{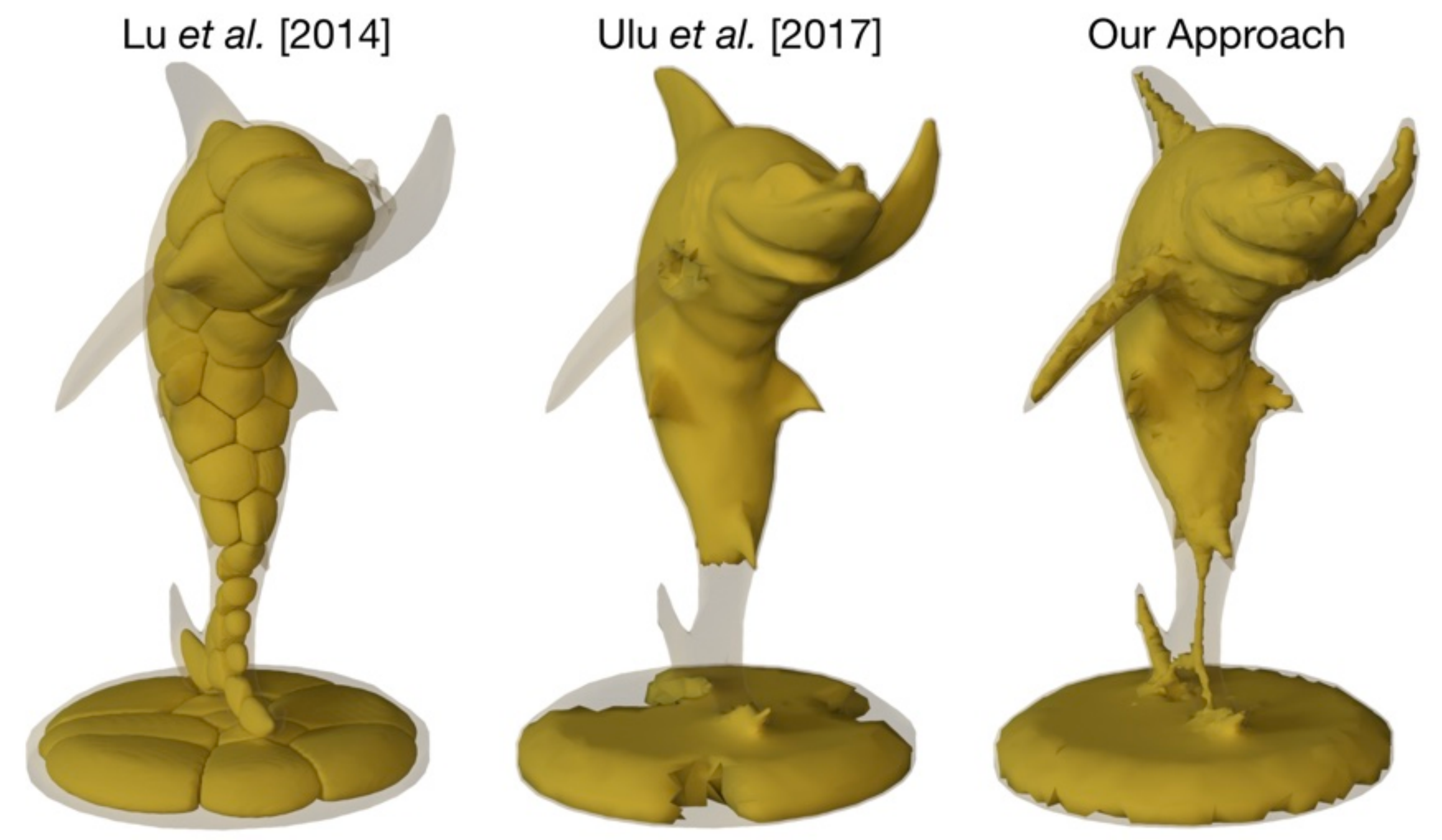}
\caption{Comparison of structures generated by Lu \etal~\cite{lu2014build}, Ulu \etal~\cite{ulu2017lightweight} and our method. Our optimization approach produces a lighter structure that has only one connected cavity inside while sustaining any possible force applied on the boundary.}
\label{fig:Comparison-FLU-BuildToLast}
\end{figure}

\paragraph*{Comparison}
In Figure~\ref{fig:Comparison-FLU-BuildToLast}, we compare our approach with the build-to-last method \cite{lu2014build} and the lightweighting method in \cite{ulu2017lightweight}. As the build-to-last method creates honeycomb-like structures, it tends to generate large number of unconnected internal voids. For the shark model, the number of such components turns out to be $48$. Therefore, unless a special technique is employed to avoid internal supports in 3D printing, post-processing could be very cumbersome and costly. Our approach, however, generates a shell structure with a single connected hole. 

In comparison to \cite{ulu2017lightweight}, our optimum result weighs $14.7\%$ less when it is optimized for $20N$ load that is allowed to be applied anywhere on the surface of the shark in normal direction. As their approach is a reduced order method, the performance is affected highly by the number of material basis being used. For $15$ material modes, the optimization can perform more global alterations than local changes. On the other hand, our method can perform more localized alterations using all the vertex temperatures and thereby hollowing out both the fins as well as the tail. 

\subsection{Validation and Performance}

\paragraph*{Fabrication}
We 3D printed our optimum shells using polyjet (OBJET Connex 350), SLS (ProMaker P2000 HT) and FDM (Strarasys F170) technologies. As our method creates a single connected void inside the object, internal supports (or excess material) can be removed by piercing a single access hole to empty the material trapped inside the shell. For models printed using SLS and polyjet, we used water jet for this purpose. However, for cases where a large amount of material can be accessed through only a small passage (such us the shark or deer head models), we observed that such an approach might not be effective. In those cases, a 3D printing technology allowing soluble support material is a more practical solution. We used soluble support in our FDM prints as an example (see accompanying video).

\begin{figure}
\centering
\includegraphics[width = 1.0\columnwidth]{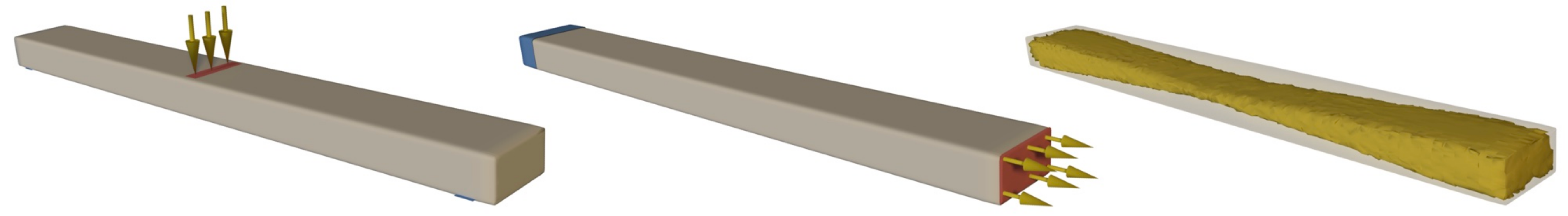}
\caption{The beam model is optimized for two different problem configurations--three-point bending (left) and tensile test (middle). Optimum result (right) satisfies the design constraints for both problem configurations concurrently by thickening the middle region.}
\label{fig:MultipleBC}
\end{figure}

\paragraph*{Physical Tests}
Figure~\ref{fig:MultipleBC} illustrates a beam model we designed to perform tests to physically evaluate the strength of the 3D printed model. The model is required to withstand $115N$ force in three-point bending and tensile loading configurations. As the stress is concentrated in the middle section under three-point bending  compared to the more uniformly distributed stresses in tensile loading case, our algorithm thickens the middle  while keeping the ends as thin as possible. 

\begin{figure}
	\includegraphics[width = \columnwidth]{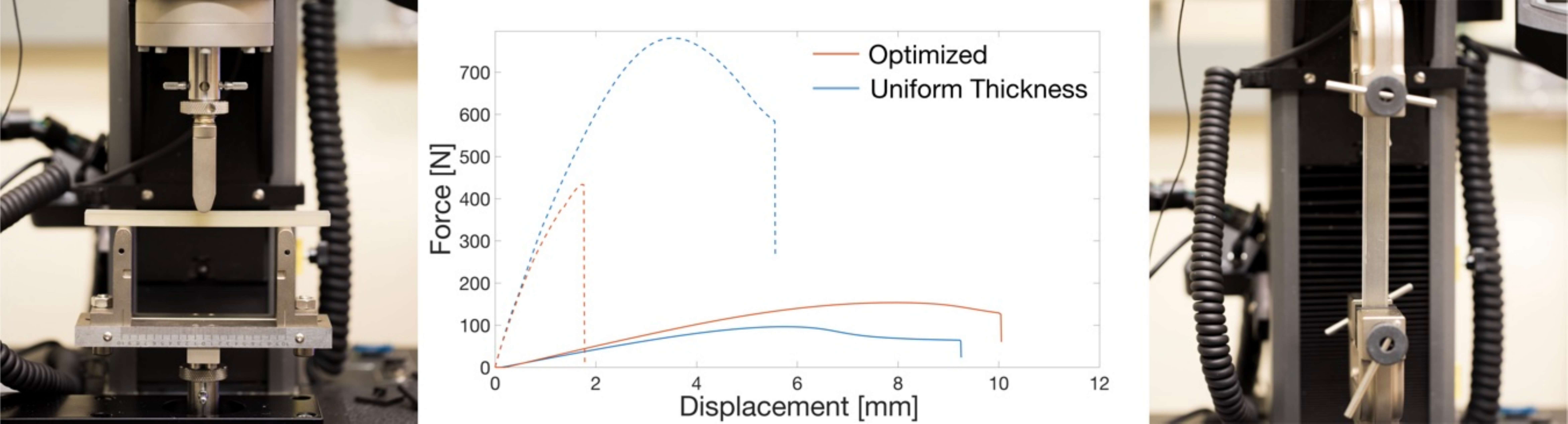}
   	\caption{Three-point bending (left) and tensile (right) tests on the optimized beam model and identical weight uniform thickness beam. The optimized model meets the design constraint of $115N$ for both problem configurations while the uniform thickness model only satisfies it for the tensile loading.}
   	\label{fig:BeamTest}
\end{figure}

To validate the optimization result, we performed three-point bending and tensile tests on our optimized beam model. We used an INSTRON universal testing machine for our tests. For comparison, we performed the same two tests on an identically weighing uniform thickness beam. Figure~\ref{fig:BeamTest} shows the test setups as well as the resulting force-displacement curves. For the uniform thickness model, we measured $96.3N$ and $780.8N$ of failure (yielding) forces for three-point bending and tensile tests, respectively. Although its tensile performance is better, the uniform thickness model fails to satisfy the design constraint of $115N$ for the three-point bending case. Our optimum result, however, can withstand $153.4N$ and $433.7N$ forces for these cases, thereby satisfying the design constraint for both. Note that, for the same mass, our method shifts the material towards the center to improve the performance for the three-point bending test and satisfy the constraint while sacrificing its tensile performance (while satisfying both imposed constraints).

\paragraph*{Performance}

Table~\ref{tab:Results} shows the performance of our algorithm. We tested our method on a computer with an 3.7GHz Intel Xeon W-2145 CPU and 32GB memory. Our tests include the optimization of various 3D models for problem configurations of different complexities. As in most of the structural optimization approaches, FEAs constitute the majority of the computational cost in our approach. Therefore, the performance is heavily affected by the size of the volumetric mesh as well as the number of problem configurations. Sea horse (large number of elements) vs. cactus (small number of elements) highlights the impact of the mesh size on the optimization time. For the same number of force configurations, computation time per iteration increases $\sim9.5\times$ on average when the number of elements is increased by $\sim2\times$. Similarly, problems with force location uncertainty (Pitcher, Spot, Mug and Shark) take longer to optimize compared to problems with a single (Cactus, Sea Horse) or a small number of problem configurations (Deer Head, Lilium, Octopus and Test beam).  However, our estimation based approach in determining $\boldsymbol{\sigma}$ and $\sigma_{cr}$ allows us to achieve $\sim5\times$ acceleration on average over a brute force approach.

\begin{figure}
	\includegraphics[width = \columnwidth]{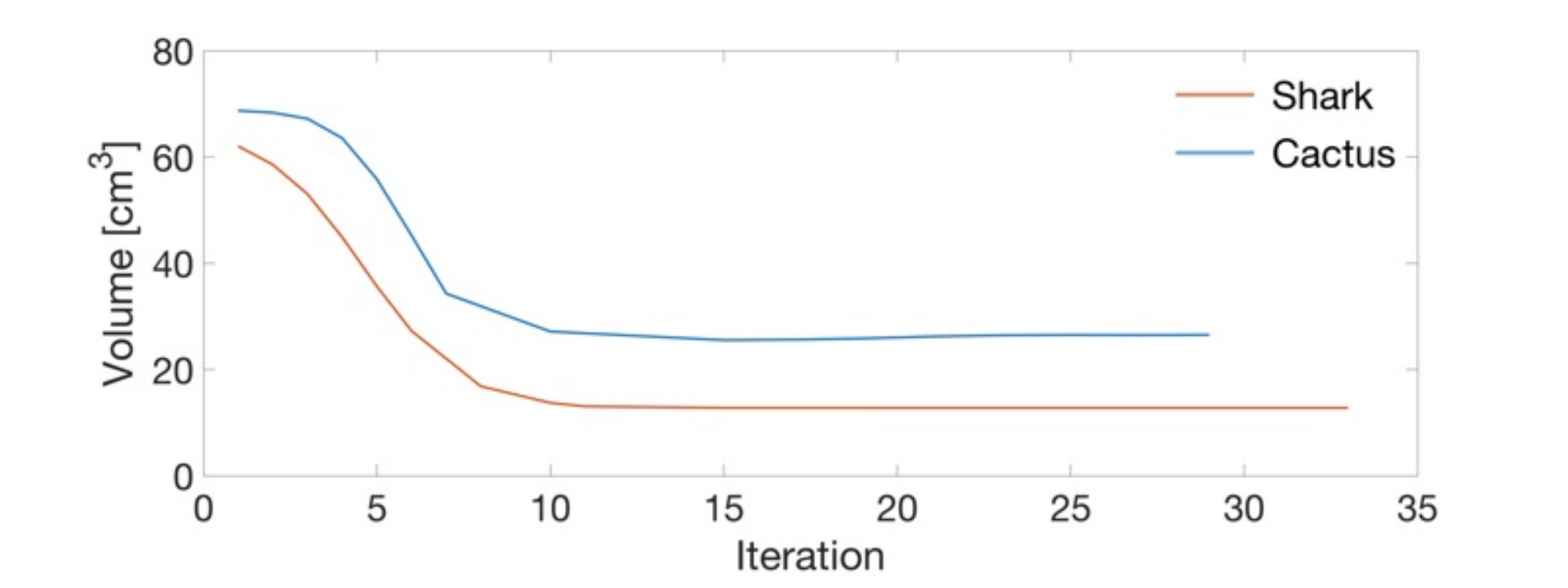}
   	\caption{Convergence of the Cactus and Shark models.}
   	\label{fig:Convergence}
\end{figure}

In all of our examples, the optimization converges under $100$ iterations. Figure~\ref{fig:Convergence} illustrates the convergence profiles for Cactus and Shark as examples of classical structural design problems with a single force configuration and problems with force location uncertainties, respectively. \red{Note that larger step sizes used at the initial stages allow the optimizer to reach reasonable solutions quickly by removing large amount of material. In later stages, however, smaller step sizes result in more local and intricate alterations requiring larger number of iterations. Although, faster convergence can be achieved by relaxing the stopping criteria, this comes at the cost of limited flexibility in optimization.}


\subsection{Limitations and Future Work}

Our algorithm can create very thin shells when required. However, the density values associated with the elements on such thin regions could be very low resulting in inaccurate stress calculations. We address this problem by enforcing a single layer of boundary elements to be solid at all times. The thickness of this layer, however, is largely dictated by the quality of the volumetric mesh. \red{Linear elasticity model may fail to predict the nonlinear buckling failure modes of shell structures. In the future, our analysis could be extended or complemented with nonlinear corotational methods to capture such failure modes more accurately.}

The temperature distribution in our algorithm depends on the geometry of the skeleton created. We found mean curvature skeletons to work well for organic shapes. However, for man-made shapes such as voluminous mechanical objects, the generated skeleton may limit the design space and restrict the quality of the resulting shell structure. For such cases, a manual adjustment tool such as a sketching interface would be beneficial to the user. For open meshes, we use a simple heuristic of projecting the input surface to a plane in creating the skeleton. Although this approach works well for smooth surfaces with low curvature, computation of the skeleton for complex open surfaces remains an open problem.

For very large boundary temperatures, our formulation is bound to create a small internal void around the skeleton (such as the spout of the pitcher model in Figure~\ref{fig:Results-FLU}). Such small voids might result in manufacturability issues in 3D printing and more importantly can create stress concentration problems. A natural extension to our approach would be to limit the minimum hole size by thickening the skeleton using morphological operations. 

\section{Conclusion}

We present a lightweight shell structure optimization method for 3D objects. We propose a heat based shape parametrization method to create shell structures with large thickness variations. With this method, we show that smooth internal surfaces can be created without self-intersections. A rapid transition between our heat based shape parametrization and the density based representation provides a practical solution to the computationally demanding design problem involving structural mechanics. Combined with the data-driven critical stress estimation approach, we demonstrate that our method can be applied to a generalized set of problems where there is uncertainty in the force locations. We evaluate the performance of our algorithm on a variety of 3D models. Our results show that significant mass reductions can be achieved by optimizing the shell thickness while ensuring that the object is structurally sound against a wide range of force configurations.

\paragraph*{Acknowledgements}
The authors would like to thank Nurcan Gecer Ulu, Walter Hsiao and Saigopal Nelaturi for insightful discussions and David Johnson for his help in 3D printing. We are grateful to the designers whose 3D models we used: kyoday (Thingiverse) for sea horse, yeg3d (Thingiverse) for deer head, xiaoxunyue2016 (Thingiverse) for octopus, HappyMoon (Thingiverse) for mug, Maninder Singh Chanay (Grabcad) for pitcher, Keenan Crane for spot, Sorkine-Hornung et al. for the cactus and Lu et al. for the shark. This work is partly funded by NCDMM America Makes Project \#4058.